\pdfoutput=1
\documentclass[
               DIV=15,
               bibliography=totoc,
               ]{scrartcl}  

\usepackage[a4paper,margin=2.25cm,bottom=2.5cm]{geometry}

\usepackage{color} 
\definecolor{linkblue}{rgb}{0,0.1,0.6}
\definecolor{citegreen}{rgb}{0,0.4,0.0}
\definecolor{linkred}{rgb}{0.8,0,0.005}
\definecolor{mailviolet}{rgb}{0.3,0,0.35}
\definecolor{tumblue}{rgb}{0,0.396,0.741}
\usepackage{url}
\usepackage{subfig}
\usepackage{amsmath}
\usepackage{amsfonts}
\usepackage{amssymb}
\usepackage{graphicx}
\usepackage{pstool} 

\usepackage{lastpage}
\usepackage{todonotes}
\usepackage{grffile}
\usepackage{placeins}
          
\usepackage[square,comma,nonamebreak,compress,numbers]{natbib}

\usepackage{pgfplots}
\usepackage{pgfplotstable}
\usepackage{filecontents}

\usepackage{ifthen}

\usepackage{shellesc}
\usepackage{tikz}
\usepackage{tikzscale}
\usepackage{tikz-3dplot}

\usetikzlibrary{spy}
\usetikzlibrary{intersections}
\usetikzlibrary{snakes,arrows,shapes}    
\usetikzlibrary{spy}
\usetikzlibrary{backgrounds}  
\usetikzlibrary{decorations}
\usetikzlibrary{decorations.markings}
\usetikzlibrary{positioning}
\usetikzlibrary{patterns}
\usetikzlibrary{calc} 
\usetikzlibrary{quotes} 
\usetikzlibrary{external} 

\usepackage{siunitx}
\usepackage{booktabs}   
\usepackage{makecell} 
\usepackage{colortbl}   
\usepackage{enumitem}

\usepackage{xspace}
\usepackage{color}     
\usepackage{setspace}   
 
\usepackage{soul}
\usepackage{soulpos}
\usepackage{ulem}
\usepackage{currfile}
\usepackage{import}

\usepackage{authoraftertitle} 
\usepackage{authblk} 

\makeatletter
\tikzoption{canvas is plane}[]{\@setOxy#1}
\def\@setOxy O(#1,#2,#3)x(#4,#5,#6)y(#7,#8,#9)%
  {\def\tikz@plane@origin{\pgfpointxyz{#1}{#2}{#3}}%
   \def\tikz@plane@x{\pgfpointxyz{#4}{#5}{#6}}%
   \def\tikz@plane@y{\pgfpointxyz{#7}{#8}{#9}}%
   \tikz@canvas@is@plane
  }
\makeatother

\makeatletter
\tikzset{spy on other/.code={%
  \pgfutil@g@addto@macro\tikz@lib@spy@collection{%
    \setbox\tikz@lib@spybox=\hbox{\pgfpicture#1\endpgfpicture}}}}
\makeatother

\usepackage[ruled,linesnumbered]{algorithm2e}

\pgfplotsset{compat=1.8}

\newcommand{\findmax}[3]{
    \pgfplotstablesort[sort key={#2},sort cmp={float >}]{\sorted}{#1}%
    \pgfplotstablegetelem{0}{#2}\of{\sorted}%
    \let #3=\pgfplotsretval%
}

\pgfplotscreateplotcyclelist{myCycleListBlackAndWhite}
{%
  {black, mark=o},
  {black, mark=square},
  {black, mark=triangle},
  {black, mark=diamond}, 
  {black, mark=pentagon}, 
  {black, mark=*},
  {black, mark=square*},
  {black, mark=triangle*},
  {black, mark=diamond*}, 
  {black, mark=pentagon*}, 
}

\definecolor{darkgreen}{rgb}{0,0.4,0} 
\definecolor{darkbrown}{rgb}{0.5, 0.396, 0.09}

\pgfplotscreateplotcyclelist{myCycleListColor}
{%
  {blue, mark=o},
  {red, mark=square},
  {darkbrown, mark=triangle},
  {darkgreen, mark=diamond},
  {black, mark=pentagon},
  {blue, mark=*},
  {red, mark=square*},
  {darkbrown, mark=triangle*},
  {darkgreen, mark=diamond*},
  {black, mark=pentagon*},
}

\tikzset{dashdot/.style={dash pattern=on .4pt off 3pt on 4pt off 3pt}} 
\pgfplotscreateplotcyclelist{myCycleLineStyleListColor}
{%
  {blue, mark=none, style=solid, smooth},
  {red, mark=none, style=dashed, smooth},
  {darkbrown, mark=none, style=dashdot, smooth},
  {darkgreen, mark=none, style=dotted, smooth},
  {black, mark=none},  
}

\pgfplotsset{every axis/.append style= 
              {
                font=\small,
                mark size=2,
                line width = 0.5,
                legend style={font=\small, mark size=3, draw=none, fill=white},
                legend cell align=left,
                cycle list name=myCycleListColor,
              }
            }
            
\pgfplotstableset
{
  every odd row/.style={before row={\rowcolor[gray]{0.9}}},
  every head row/.style=
  {
    before row=
    {%
      \toprule
    },
    after row=\midrule
  },
  every last row/.style=
  {
    after row=\bottomrule
  }
}

\newif\ifdrawboundingbox

\usetikzlibrary{external} 
\tikzset{external/system call={lualatex \tikzexternalcheckshellescape
-halt-on-error -interaction=batchmode -jobname "\image" "\texsource"}} 

\pgfkeys
{ 
  /pgf/number format/.cd,
  fixed,
  set thousands separator={\,},
  1000 sep in fractionals,
}

\tikzset{external/up to date check={md5}}

\usepackage{letltxmacro}
\LetLtxMacro{\oldmissingfigure}{\missingfigure}
\renewcommand{\missingfigure}[2][]{\tikzexternaldisable\oldmissingfigure[{#1}]{#2}\tikzexternalenable}

\LetLtxMacro{\oldtodo}{\todo}
\renewcommand{\todo}[2][]{\tikzexternaldisable\oldtodo[#1]{#2}\tikzexternalenable}

\newcommand\setHfillkern[1]{\def\Hfillkern{#1}}
\setHfillkern{-.25\textwidth}

\newcolumntype{C}[1]{>{\centering\arraybackslash}m{#1}}
\newcolumntype{R}[1]{>{\raggedright\arraybackslash}m{#1}}
\newcolumntype{L}[1]{>{\raggedleft\arraybackslash}m{#1}}

\title{\vspace{0cm}Direct structural analysis of domains defined by point clouds}

\author[1]{L\'{a}szl\'{o} Kudela}
\author[1]{Stefan Kollmannsberger}
\author[2]{Umut Almac}
\author[1,3]{Ernst Rank}

 \affil[1]{Chair for Computation in Engineering,
 Technical University of Munich,\authorcr
  Arcisstr. 21, 80333 M\"unchen, Germany}

\affil[2]{Istanbul Technical University, Faculty of Architecture, Turkey}
 
 \affil[3]{Institute for Advanced Study,
 Technical University of Munich, \authorcr
 Lichtenbergstr. 2a, 85748 Garching, Germany}

\newcommand{\publicationDate}{\today}

\usepackage{fancyhdr}
\date{}
\fancypagestyle{plain}{%
  \fancyhf{}%
  \fancyfoot[L]{
  \footnotesize{Preprint submitted to \textit{
    CMAME}  \hfill
  \publicationDate
  \\
  }} }

\begin{document}    
\normalem
\maketitle  
  
\normalfont\fontsize{11}{13}\selectfont
\vspace{-1.5cm}  
\hrule 
\section*{Abstract}
This contribution presents a method that aims at the numerical analysis of solids represented by oriented point clouds.
The proposed approach is based on the Finite Cell Method, a high-order immersed boundary technique that computes on a regular background grid of finite elements and requires only inside-outside information from the geometric model.
It is shown that oriented point clouds provide sufficient information for these point-membership classifications.
Further, we address a tessellation-free formulation of contour integrals that allows to apply Neumann boundary conditions on point clouds without having to recover the underlying surface.
Two-dimensional linear elastic benchmark examples demonstrate that the method is able to provide the same accuracy as those computed with conventional, continuous surface descriptions, because the associated error can be controlled by the density of the cloud.
Three-dimensional examples computed on point clouds of historical structures show how the method can be employed to establish seamless connections between digital shape measurement techniques and numerical analyses.

\vspace{0.25cm}
\noindent \textit{Keywords:} finite cell method, point clouds, image-based structural analysis
 
\vspace{0.25cm}


\section{Introduction}
\label{sec:introduction}
It is well known in the computational mechanics community that transferring a CAD model into an analysis-suitable finite element mesh may account for as much as 80\% of the entire analysis time~\cite{Cottrell:2009}.
Recent years' research efforts aiming at circumventing this bottleneck have resulted in numerous alternative approaches, leading to significant progress in the quest of establishing seamless connections between geometric modeling and finite element analysis.

However, in some applications of the FEM, the geometries of interest are not directly available in form of CAD models.
Typically, this situation arises in the context of biomechanical simulations where models are recorded by means of medical imaging techniques, such as CT scans.
As standard finite elements require a CAD model to start with, these methods require special algorithms to recover a geometric model and eventually a finite element mesh from the imaging data -- see e.g.~\cite{Zhang:2013} for a conceptual overview of these multi-step pipelines. 

Volumetric imaging is not always the most feasible approach to record the shape of physical structures.
It is especially large objects that do not allow for a cost-effective application of CT scanning.
Nonetheless, it can be of importance to be able to compute the structural behavior of large objects -- e.g. in the field of cultural heritage preservation, as there are often no digital CAD models available for historical structures.
Moreover, even if there are schematic drawings, the shape of the object may differ from them, especially if the structure is exposed to damaging effects such as erosion, floods, earthquakes, or wars.
In these cases, other shape measurement techniques need to be employed.
The two most popular methods for this purpose are terrestrial laser scanning and close range photogrammetry-based reconstructions. 
Especially photogrammetry has gained a lot of attention recently, due to the inexpensiveness of the required equipment and because of the rapid development of the computational resources as well as the associated algorithms that allow for efficient, almost real-time reconstructions~\cite{Kolev:2012}.

The methods of laser scanning and photogrammetry both reproduce the shape of the geometry of interest in the form of point clouds: a set of points representing the surface of the object.
Such point clouds are not directly suited for numerical analysis.
In order to transform the recorded data into an analysis-suitable model, it needs to pass through several stages, similar to the necessary procedure for models stemming from volumetric imaging.

\begin{figure}[!tbp]
  \centering
  \psfragfig[width=1.0\textwidth]{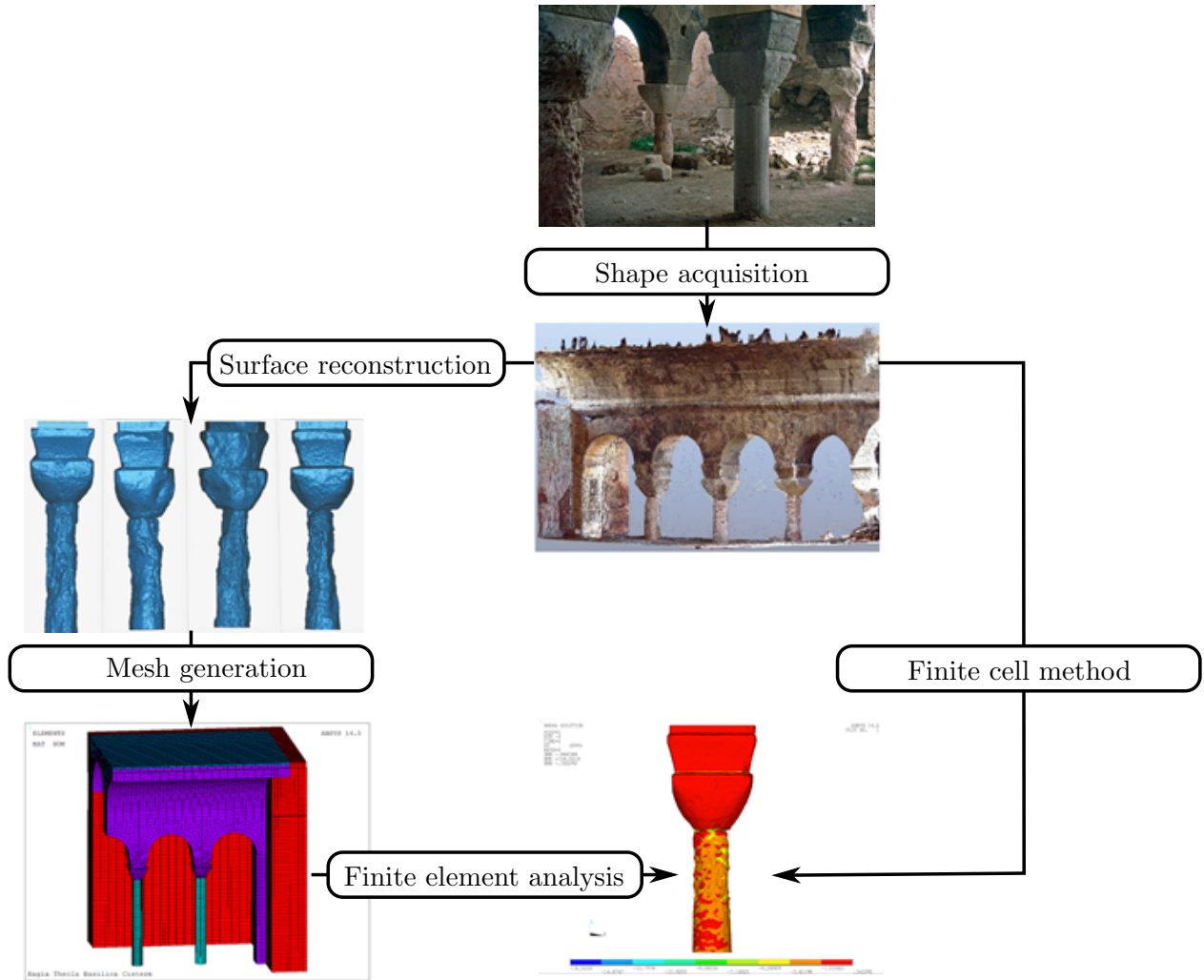}
  {
    \psfrag{sa}[c]{Shape acquisition}
    \psfrag{gr}[c]{Surface reconstruction}
    \psfrag{mg}[c]{Mesh generation}
    \psfrag{fea}[c]{Finite element analysis}
    \psfrag{pcfcm}[c]{Finite cell method}
  }
  \caption{\textit{From point clouds to simulations: standard pipeline (left) and proposed pipeline (right)}}
  \label{fig:stairwayToHeaven}
\end{figure}

Usually, these \textit{measurement-to-analysis} procedures are characterized by the following main steps (Figure~\ref{fig:stairwayToHeaven}):
\begin{enumerate}
  \item \textit{Shape acquisition}\\
    A 3D shape measurement technique is employed to capture the shape of the domain of interest, resulting in a point cloud representing the surface of the object.
  \item \textit{Surface reconstruction}\\
    A geometric model is derived from the point cloud information using geometric segmentation and surface fitting methods. The resulting model is stored using standardized geometric representation techniques, such as STL, STEP, or IGES files.
  \item \textit{Mesh generation}\\
    The CAD model from the previous step is discretized into a finite element mesh. 
  \item \textit{Finite Element Analysis}\\
    The mesh is handed over to a finite element solver together with the corresponding material properties and structural constraints.
\end{enumerate}

Numerous applications implement the steps above -- see e.g. \cite{Kalisperakis:2015,Borri:2006,Riveiro:2011,Almac:2016,Castellazzi:2015} for examples in the preservation of historical structures, or~\cite{Kudela:2017} for an application in the context of biomechanical experiments.

Research in different fields of computational science and engineering has resulted in well-established approaches that allow to perform these steps one-by-one.
Still, their deep integration into a seamless chain is not trivial, as it requires the interplay of various algorithms.
Other than the problems inherent to the data transfer between different implementations, an even bigger challenge is posed by generating a finite element mesh from the geometric model reconstructed in the second step.
The fine details recovered by modern surface reconstruction algorithms (e.g.\cite{Kazhdan:2006,Kazhdan:2013,Calakli:2011}) are not necessarily the details that need to be carried over to a finite element mesh, where the process of refinement is usually governed by the physics of the problem rather than aesthetic aspects.
While a geometric defeaturing step may be applied to remove physically uninteresting details, manipulating the geometry carries the danger of introducing flaws in the geometric model, resulting in an invalid, ''dirty'' geometry that cannot be meshed directly~\cite{Beall:2003,Wassermann:2018}.

One method that aims to avoid the difficult task of mesh generation is the Finite Cell Method (FCM), introduced in~\cite{Parvizian:2007}. 
The FCM is based on the combination of immersed boundary methods and high-order finite element basis functions used in p-FEM~\cite{Szabo:2004} or Isogeometric Analysis~\cite{Cottrell:2009}.

Instead of generating a boundary-conforming discretization, the FCM extends the physical domain of interest by a so-called fictitious domain in such a way that their union forms a simple bounding box that can be meshed easily. 
To stay consistent with respect to the original problem, the material parameters in the fictitious domain are penalized by a small factor $\alpha$.
The introduction of $\alpha$ shifts the analysis effort from mesh generation to numerical integration.
The most notable advantages of the FCM are the drastically reduced engineering efforts for preprocessing, the almost costless meshing, and the high accuracy and efficiency of the computation.

In its simplest implementation, the only information that the FCM needs from a geometric model is the inside-outside state -- the question whether a given point in space lies in the physical or the fictitious part of the domain.
There are numerous geometric representations that are suitable to provide such \textit{point membership tests} and that have been shown to work well in combination with the FCM, ranging from simple shapes provided by constructive solid geometry~\cite{Wassermann:2017} to models as complex as metal foams~\cite{Heinze:2018}.

In this paper, we will demonstrate that a direct analysis of geometries described by oriented point clouds is possible. 
To this end, we will combine the finite cell method with geometries that are represented by oriented point clouds.
The members of the point cloud and the vectors associated to them provide enough information for point membership tests, allowing for structural analyses of objects directly on their cloud representation.
This way, the tedious tasks of recovering a geometric model and generating a boundary conforming mesh can be avoided, allowing for significant simplifications in the measurement-to-analysis pipeline.

\section{The finite cell method combined with oriented point clouds}
This section summarizes the basics of the finite cell method as well as the concept of performing inside-outside tests on oriented point clouds. 
The description is limited only to the necessary minimum for the context of this paper. 
For further details, refer to~\cite{Duster:2017}.

\subsection{The finite cell method}
\label{ssec:fcm}
\begin{figure}
  \centering
  \psfragfig[width=\textwidth]{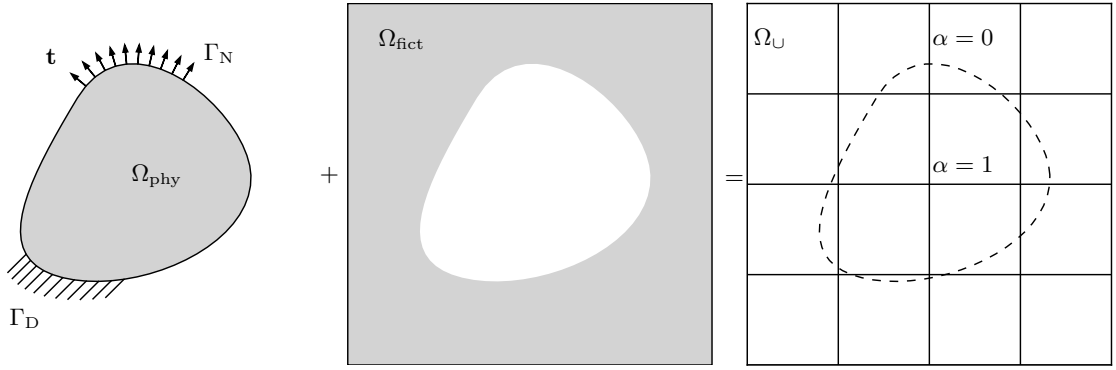}
  {
    \footnotesize
    \psfrag{p}{$\Omega_{\text{phy}}$}
    \psfrag{f}{$\Omega_{\text{fict}}$}
    \psfrag{p1}{$+$}
    \psfrag{e}{$=$}
    \psfrag{a0}{$\alpha=0$}
    \psfrag{a1}{$\alpha=1$}
    \psfrag{ou}{$\Omega_{\cup}$}
    \psfrag{t}{$\mathbf{t}$}
    \psfrag{gn}{$\Gamma_\text{N}$}
    \psfrag{gd}{$\Gamma_\text{D}$} 
  }
\caption{\textit{The core concept of the FCM. The physical domain $\Omega_{\text{phy}}$ is extended by the fictitious domain $\Omega_{\text{fict}}$. Their union, the embedding domain $\Omega_{\cup}$ can be meshed easily. The influence of the fictitious domain is penalized by the scaling factor $\alpha$.}}
\label{fig:fcmIdea}
\end{figure}

Figure~\ref{fig:fcmIdea} illustrates the core idea of the FCM. The boundaries of the physical domain of interest $\Omega_{\text{phy}}$ are extended by a fictitious part $\Omega_{\text{fict}}$. Their union $\Omega_{\text{phy}} \cup \Omega_{\text{fict}}$ forms the embedding domain $\Omega_{\cup}$. As $\Omega_{\cup}$ possesses a simple, box-like geometry, it can easily be meshed into a structured grid of rectangular finite elements in 2D and cuboids in 3D.

Similar to standard finite elements, the FCM is derived from the principle of virtual work~\cite{Hughes:2000}:
\begin{equation}
  \begin{aligned}
  \label{eq:virtualWork}
    \delta W(\mathbf{u},\delta\mathbf{u}) = \int\displaylimits_{\Omega}\boldsymbol{\sigma}:\left(\nabla_{\text{sym}}\delta\mathbf{u}\right)\text{d}V &- 
    \int\displaylimits_{\Omega_{\text{phy}}}\delta\mathbf{u}\cdot\mathbf{b}\text{d}V -
    \int\displaylimits_{\Gamma_{\text{N}}}\delta\mathbf{u}\cdot\mathbf{t}\text{d}A = 0 \\
    \boldsymbol{u}&=\boldsymbol{u}_0 \qquad \forall\boldsymbol{x}\in\Gamma_{\text{D}},
  \end{aligned}
\end{equation}
where $\boldsymbol{\sigma},\boldsymbol{b},\boldsymbol{u},\delta\boldsymbol{u}\text{ and }\nabla_{\text{sym}}$ denote the Cauchy stress tensor, the body forces, the displacement vector, the test function, and the symmetric part of the gradient, respectively. 
Prescribed displacements $\boldsymbol{u}_0$ are defined on the boundary~$\Gamma_{\text{D}}$, while the traction vector $\boldsymbol{t}$ specifies the Neumann boundary conditions on $\Gamma_{\text{N}}$, such that $\Gamma_{\text{D}} \cap \Gamma_{\text{N}} = \varnothing$.

The stresses and strains are related through the constitutive tensor $\mathbf{C}$:
\begin{equation}
\boldsymbol{\sigma}=\alpha \mathbf{C} : \boldsymbol{\varepsilon},
\end{equation}
where $\alpha$ is an indicator function defined as:
\begin{equation}
\label{eq:penalty}
\alpha(\boldsymbol{x})=\begin{cases}
1 & \forall\boldsymbol{x} \in \Omega_{phy} \\
10^{-q} & \forall\boldsymbol{x} \in \Omega_{fict}.
\end{cases}
\end{equation}
To avoid ill-conditioning of the resulting equation system, the value of $q$ is chosen between $6$ and $12$ for practical applications.

Homogeneous Neumann boundary conditions are automatically satisfied. Nonhomogeneous Neumann boundary conditions can be realized by evaluating the contour integral over $\Gamma_\text{N}$ in Equation~\ref{eq:virtualWork}. 
However, in contrast to standard finite elements, Dirichlet constraints cannot be applied directly, as $\Gamma_{\text{D}}$ usually does not coincide with the element boundaries. 
Therefore, these boundary conditions are often formulated in the weak sense, e.g. using the penalty method or Nitsche's method, e.g.~\cite{Ruess:2013}.

The unknown quantities $\delta\mathbf{u}$ and $\mathbf{u}$ are discretized by a linear combination of shape functions $\text{N}_i$ with unknown coefficients $\mathbf{u}_i$:
\begin{equation}
\label{eq:discreteDisplacement}
\mathbf{u}=\sum_i\text{N}_i\mathbf{u}_i \text{ ; } \delta\mathbf{u}=\sum_i\text{N}_i\delta\mathbf{u}_i.
\end{equation}
The finite cell method uses shape functions $\boldsymbol{N}_i$ of higher order. 
Popular choices are either the integrated Legendre polynomials known from p-FEM~\cite{Duster:2017} or spline-based shape functions used in isogeometric analysis~\cite{Cottrell:2009}.

Following the standard Bubnov-Galerkin approach~\cite{Zienkiewicz:1977,Hughes:2000}, substituting Equation~\ref{eq:discreteDisplacement} into Equation~\ref{eq:virtualWork} leads to the discrete finite cell representation
\begin{equation}
\mathbf{K}\mathbf{u}=\mathbf{f},
\end{equation}
where $\mathbf{K}$,$\mathbf{u}$,$\mathbf{f}$ denotes the stiffness matrix, the unknown displacement coefficients, and the load vector, respectively.
The stiffness matrix results from a proper assembly of the element stiffness matrices:
\begin{equation}
  \label{eq:elementStiffness}
  \mathbf{k}^e=\int\displaylimits_{\Omega^{\text{e}}}\left[\mathbf{L}\mathbf{N}^e\right]^T \alpha(\boldsymbol{x}) C \left[\mathbf{L}\mathbf{N}^e\right] \text{d}\Omega^\text{e},
\end{equation}
where $\mathbf{L}$ is the standard strain-displacement operator, $\mathbf{N}^e$ is the matrix of shape functions associated to the element.
In the context of the FCM, the elements in the background mesh are often referred to as \textit{cells}.

Due to the introduction of the scaling factor $\alpha(\boldsymbol{x})$ in Equation~\ref{eq:elementStiffness}, the integrand becomes discontinuous.
Standard quadrature schemes of the finite element method, such as the Gauss-Legendre rule, lose their precision in the presence of this discontinuity.
In order to reduce the associated integration error, the above integral is usually evaluated by means of specially constructed quadrature schemes.
The most popular method is based on a composed Gaussian quadrature rule combined with a recursive subdivision of the elements cut by the boundary of the physical domain.
\begin{figure}[!bhtp]
  \centering
  \psfragfig[width=1.0\textwidth]{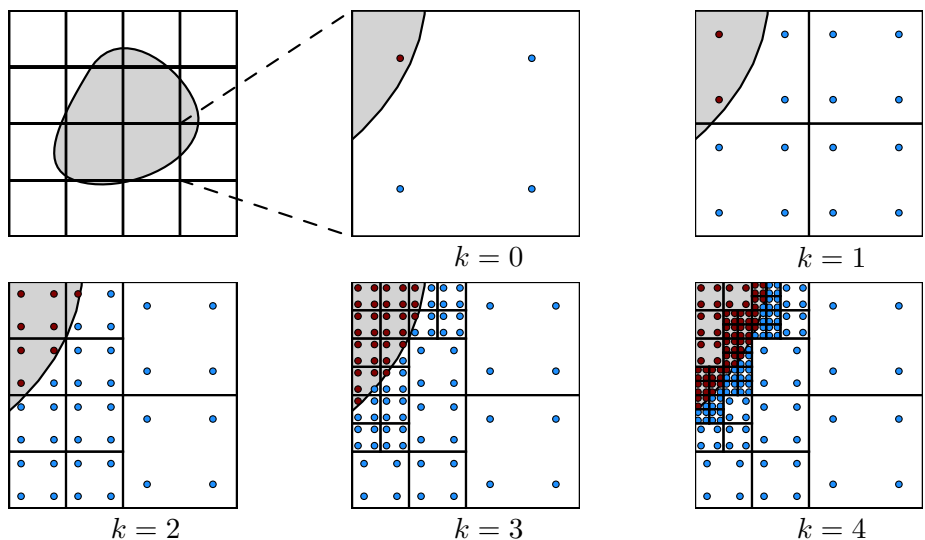} 
  {
    \psfrag{k0}{$k=0$}
    \psfrag{k1}{$k=1$}
    \psfrag{k2}{$k=2$}
    \psfrag{k3}{$k=3$}
    \psfrag{k4}{$k=4$}
  }
  \caption{\textit{Spacetree-based integration domains for different values of maximum subdivision depth~$k$. Red dots represent integration points that lie in $\Omega_{\text{phy}}$, blue dots are in $\Omega_{\text{fict}}$.}}
  \label{fig:quadtree}
\end{figure}
In this process, every intersected element is subdivided into equal subcells until a pre-defined depth $k$ is reached.
Quadrature points are then distributed on the domains of the leaf cells of this integration mesh.
An example for the two-dimensional case is depicted in Figure~\ref{fig:quadtree}. 

In addition to spacetree-based schemes, numerous alternative approaches have been developed for the purpose of integrating through the discontinuous jump. 
These schemes reduce the integration error either by decomposing the integration domain into boundary-conforming integration cells~\cite{Kudela:2015,Kudela:2016}, or by modifying the quadrature weights, e.g. using moment-fitting equations~\cite{Hubrich:2018}.

The indicator function in Equation~\ref{eq:penalty} has to be evaluted for every quadrature point.
This requires the geometric model that represents $\Omega_{\text{phy}}$ to provide point-membership tests: given a quadrature point, does this point belong to $\Omega_{\text{phy}}$ or not? 

\begin{figure}[!htbp]
  \centering
  \includegraphics[width=0.9\textwidth]{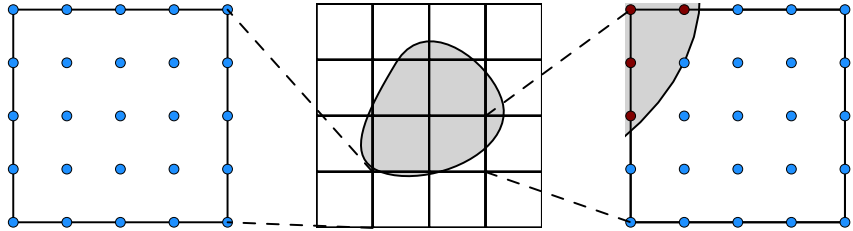}
  \caption{\textit{The process of determining whether a cell is cut by the interface $\partial\Omega_{\text{phy}}$. Seed points are distributed on every cell of the finite cell mesh, and their inside-outside state is evaluated. If there is a pair of points with differing state, the cell is identified as a cut cell. Red and blue dots represent seed points lying in $\Omega_{\text{phy}}$ or $\Omega_{\text{fict}}$, respectively. The cell on the right side is cut, while the one the left side is not.}}
  \label{fig:potatoIsCellCut}
\end{figure}

Point membership tests can also be used to control the refinement process of the spacetree-based integration.
To determine whether a cell is cut by the boundary of the physical domain, a set of test points is distributed inside the cell.
Then, the inside-outside state of every test point is evaluated.
The cell is definitely cut by the boundary of the physical domain if there is at least one pair of test points in the cell with changing inside-outside state.
This idea is depicted in Figure~\ref{fig:potatoIsCellCut}.

The consequence of the above considerations is that, in the standard case, the only information that a geometric model needs to provide for the FCM is a robust point membership classification. 
Many geometric representations are able to answer such queries and have been successfully applied in combination with the FCM. 
Examples include voxel models from CT-scans~\cite{Ruess:2012}, constructive solid geometries~\cite{Wassermann:2017}, boundary representations~\cite{Kudela:2016} and STL descriptions~\cite{Elhaddad:2015}.

\subsection{Inside-outside testing on point clouds}
In point-cloud-based simulations, the domain $\Omega_{\text{phy}}$ is represented by a set of sample points $\mathbf{p}_i$ and their associated normal vectors $\mathbf{n}_i$. 
If no outliers are present, the set of pairs $S=\left\{\mathbf{p}_i,\mathbf{n}_i\right\}$ represent a discrete sampling of the boundary $\partial\Omega_{\text{phy}}$ of the domain.

\begin{figure}
  \centering
  \psfragfig[width=\textwidth]{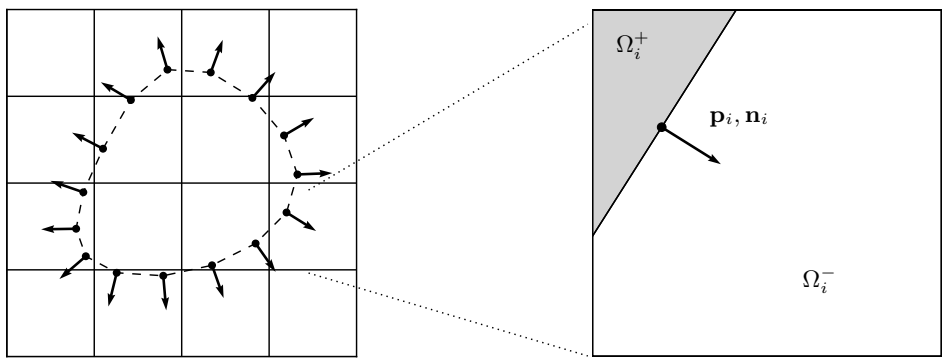}{
    \footnotesize
    \psfrag{op}{$\Omega_{i}^+$}
    \psfrag{om}{$\Omega_{i}^-$}
    \psfrag{p}{$\mathbf{p}_i,\mathbf{n}_i$}
  }
  \caption{\textit{Point membership classification on oriented point clouds. The domain is represented by a set of points $\mathbf{p}_i$ and associated normals $\mathbf{n}_i$. Every such pair locally separates the space along a hyperplane into two half-spaces: $\Omega_{i}^-$ and $\Omega_{i}^+$.}}
  \label{fig:pointCloudInsideOutside}
\end{figure}

Each element in $S$ defines a hyperplane that separates the space in two half spaces: the open half-space $\Omega_{i}^-$ on the side of the hyperplane where the normal vector $\mathbf{n}_i$ points, and the closed half-space $\Omega_{i}^+$ on the other side. 
This concept is depicted in Figure~\ref{fig:pointCloudInsideOutside}.
For every $\mathbf{x} \in \Omega_{i}^+$, the following holds:
\begin{equation}
  \label{eq:pointCloudInsideOutside}
  \left(\mathbf{p}_i - \mathbf{x}\right)\cdot \mathbf{n}_i \geq 0.
\end{equation}
Therefore, a simple way to estimate whether a quadrature point $\mathbf{q}$ lies inside or outside the domain is to find the $\mathbf{p}_i$ and the associated $\mathbf{n}_i$ in $S$ that lies closest to $\mathbf{q}$, and to evaluate the scalar product of Equation~\ref{eq:pointCloudInsideOutside}. 
While the heuristic nature of this approach may give the impression that it only works for simple shapes, our practical examples in Section~\ref{ssec:3dexamples} on more complex geometries show that the recovered indicator function is suitable to perform a finite cell analysis in these cases as well.
The algorithm requires an efficient nearest neighbor query. 
In our examples, we use the k-d tree implementation of the C++ library \textit{nanoflann}~\cite{Blanco:2014}, while the clouds themselves are represented by the data structures of the Point Cloud Library~\cite{Rusu:2011}.
The point membership classification method is summarized in Algorithm~\ref{alg:ioStateInCloud}.
\begin{algorithm}[h!]
  \SetAlgoNoLine
   \caption{\textit{Point membership test for oriented point clouds}}
  \label{alg:ioStateInCloud}
  \underline{function isPointInside} $\left(\mathbf{q},S\right)$ \;
  \SetKwInOut{Input}{Input}
  \SetKwInOut{Output}{Output}
  \Input{Quadrature point $\mathbf{q}$ and oriented point cloud $S=\left\{\mathbf{p}_i,\mathbf{n}_i\right\}$}
  \Output{Boolean true if $\mathbf{q}$ lies inside the domain represented by $S$, false otherwise}
  $\mathbf{p}_i,\mathbf{n}_i$ = getClosestPointInCloud$\left(\mathbf{q},S\right)$\;
  $\mathbf{v}$ = $\mathbf{p}_i-\mathbf{q}$\;
  $d$ = $\mathbf{v}\cdot\mathbf{n}_i$\;
  \If{$d\geq0$}{return true\;}
  return false;
\end{algorithm}
\subsection{Point-based surfaces and Neumann boundary conditions}
\label{ssec:pointCloudBC}
To apply boundary conditions in the weak sense, the contour integral in Equation~\ref{eq:virtualWork} needs to be evaluated.
For surface models, this is a relatively easy procedure, as they usually possess (or can be converted into) tessellations.
Then, the integral over $\Gamma_{\text{N}}$ is computed as the sum of the integrals over the individual simplices in the tessellation.

However, for point-based geometries, no such tessellations exist.
Although there are methods that are able to recover a triangulation from point cloud descriptions, their application would require to perform the same steps as the standard steps of the measurement-to-analysis pipeline in~Section~\ref{sec:introduction}.

\begin{figure}[h!]
  \centering
  \includegraphics[width=0.5\textwidth]{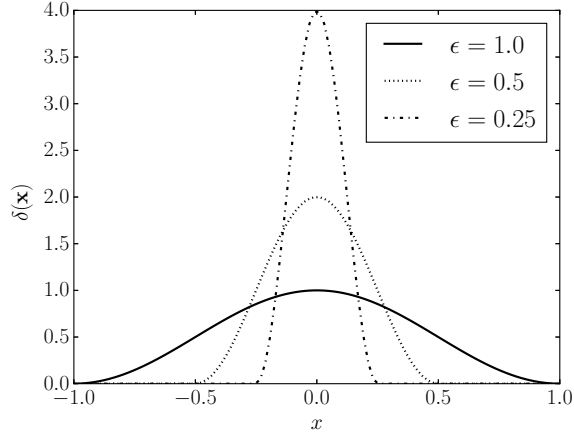}
  \caption{\textit{Regularized Dirac delta functions for different length scales.}}
  \label{fig:regularizedDiracDelta}
\end{figure}

What is needed is an alternative formulation that allows for applying boundary conditions directly on point cloud-based surface representations. 
One possible solution to this challenge is to convert the contour integral into a domain integral by using the sifting property of the Dirac delta distribution~\cite{Kublik:2013}:
\begin{equation}
  \label{eq:diracDeltaExact}
  \int\displaylimits_{\Gamma} f(\boldsymbol{x}) d\Gamma = \int\displaylimits_{\Omega} f(\boldsymbol{x}) \delta(\boldsymbol{x}) d\Omega,
\end{equation}
with
\begin{equation}
  \delta(\boldsymbol{x})=
  \begin{cases}
    \infty & \forall \boldsymbol{x} \in \Gamma \\
    0 & \text{otherwise}.
  \end{cases}
\end{equation}
In numerical applications, the regularized variant of the Dirac delta distribution is employed.
There are different choices available for the regularization, see e.g.~\cite{Engquist:2005,Lee:2012}.
In our examples, we employ the following 1D formulation:
\begin{equation}
  \label{eq:regularizedDiracDelta}
  \delta(x) \approx \delta_{\epsilon}(x) = 
  \begin{cases}
    \frac{1}{2\epsilon}\left(1+\cos\left(\frac{\pi x}{\epsilon}\right)\right) \quad &\text{if} \left| x \right| \leq \epsilon, \\
    0 & \text{otherwise},
  \end{cases}
\end{equation}
where $\epsilon$ is a length scale parameter that controls the width of the regularization.
Figure~\ref{fig:regularizedDiracDelta} depicts $\delta_{\epsilon}$ for different choices of $\epsilon$.

\begin{figure}
  \centering
  \psfragfig[width=0.8\textwidth]{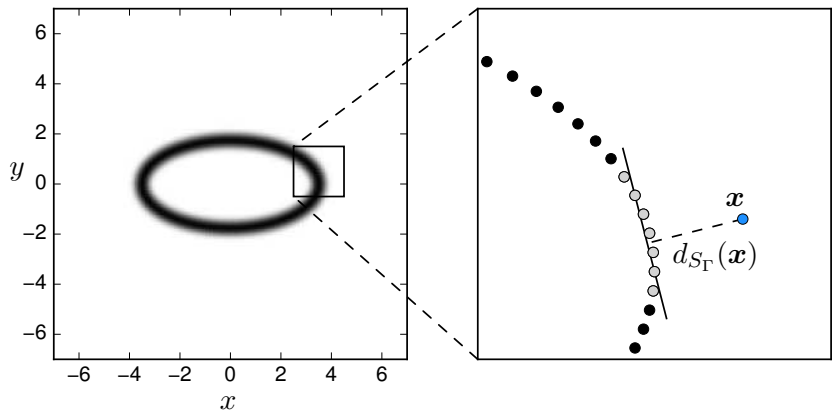}
  {
    \psfrag{xa}{$x$}
    \psfrag{y}{$y$}
    \psfrag{xq}{$\boldsymbol{x}$}
    \psfrag{d}{$d_{S_{\Gamma}}(\boldsymbol{x})$}
  }
  \caption{\textit{Computing the distance $d_{S_{\Gamma}}(\boldsymbol{x})$ towards the point set $S_{\Gamma}$. For a given query point $\boldsymbol{x}$, the nearest point $\boldsymbol{p}^*$ and its n-neighborhood $\left\{\boldsymbol{p}_i\right\},i=1...n$ is found, depicted as gray dots. Here, $n=6$. Then, the distance towards the planar approximant on $\left\{\boldsymbol{p}_i\right\}$ is computed. The resulting delta field is shown on the left side.}}
  \label{fig:pcaDistance}
\end{figure}

To extend the delta function to more dimensions, the distance function $d_{\Gamma}(\boldsymbol{x}): \mathbb{R}^n \rightarrow \mathbb{R}$ is needed, which, for a given $\boldsymbol{x}\in\mathbb{R}^n$, returns the distance of $\boldsymbol{x}$ to the contour $\Gamma$. 
Then, the multi-dimensional regularized delta function, associated to the contour $\Gamma$, can be written as:
\begin{equation}
  \label{eq:multiDimensionalDelta}
  \delta_{\Gamma}(\boldsymbol{x}) = \delta_{\epsilon}(d_{\Gamma}(\boldsymbol{x})).
\end{equation}
In the point-cloud setting, $\Gamma$ is represented by the point set $S_{\Gamma} \subset S$.
Instead of computing the distance to closest point $\boldsymbol{p}_i$, we compute the planar approximation of the n-neighborhood of $\boldsymbol{p}_i$ based on principal component analysis.
Then, we evaluate the distance toward this approximant, as depicted in Figure~\ref{fig:pcaDistance}.

Finally, the contour integral for the Neumann boundary condition in Equation~\ref{eq:virtualWork} is formulated as:
\begin{equation}
  \label{eq:discreteNeumann}
  \int\displaylimits_{\Gamma_{\text{N}}}\delta\mathbf{u}\cdot\mathbf{t}\text{d}A \approx 
  \int\displaylimits_{\Omega_{\cup}}\delta_{\epsilon}\left(d_{S_{\Gamma}}\left(\boldsymbol{x}\right)\right)\left(\delta\mathbf{u}\cdot\mathbf{t}\right)\text{d}\Omega 
\end{equation}
It is noted here that the transformation of boundary condition integrals into a domain integral using the delta function is an already existing concept in the context of FCM. 
Similar to point-cloud descriptions, phase-field models do not possess tessellations either. 
In this case, the approach outlined in this section can be applied to compute Neumann boundary conditions as well as Dirichlet boundary conditions formulated in the weak sense.
For more details, refer to~\cite{Stoter:2017,Nguyen:2018}.

\subsection{Outliers}
Sometimes, the cloud acquired by the shape measurement process carries outliers. As defined in~\cite{Hawkins:1980}, outliers are \textit{,,observations that deviate so much from other observations as to arouse suspicion that it was generated by a different mechanism.''}
The detection and treatment of outliers -- an aspect that lies beyond the scope of this article -- is a well-studied problem in the literature, see e.g. the methods in~\cite{Hawkins:1980,Lu:2003,Sotoodeh:2006,Wolff:2016,Nurunnabi:2015,Zhang:2009,Schall:2015}.
These algorithms can be applied in a pre-processing step in order to recover a clean cloud where the majority of the outliers have been removed.
The cleaning procedure is an essential step also in the traditional measurement-to-analysis pipeline, as most of the geometry recovery algorithms rely on a clean input cloud.

\begin{figure}[!tbp]
  \centering
  \subfloat[Inside-outside testing using a single nearest neighbor \label{fig:outlier1} ]{
    \includegraphics[width=0.5\textwidth]{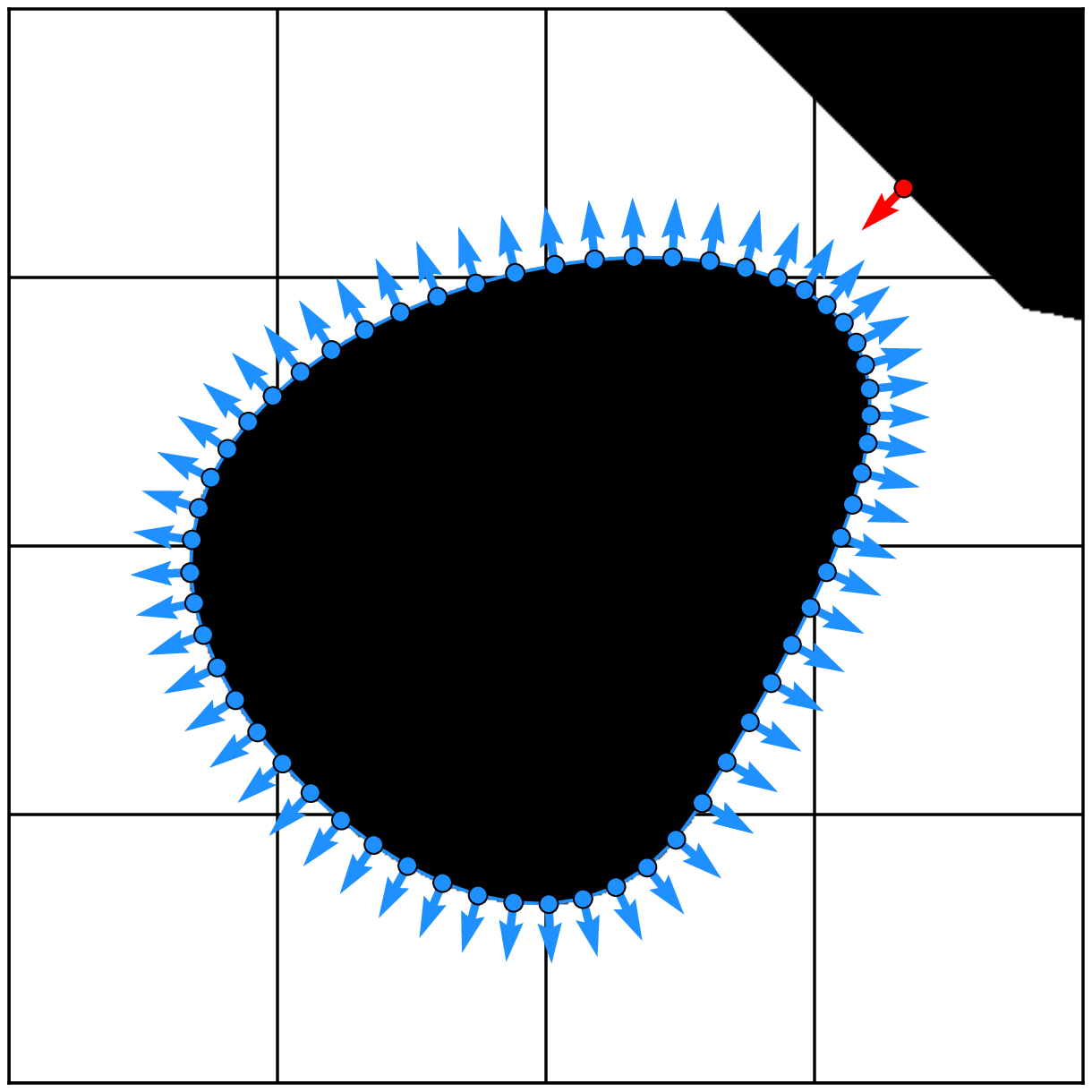} 
  }%
  \subfloat[Inside-outside testing using 3 nearest neighbors \label{fig:outlier2}]{
    \includegraphics[width=0.5\textwidth]{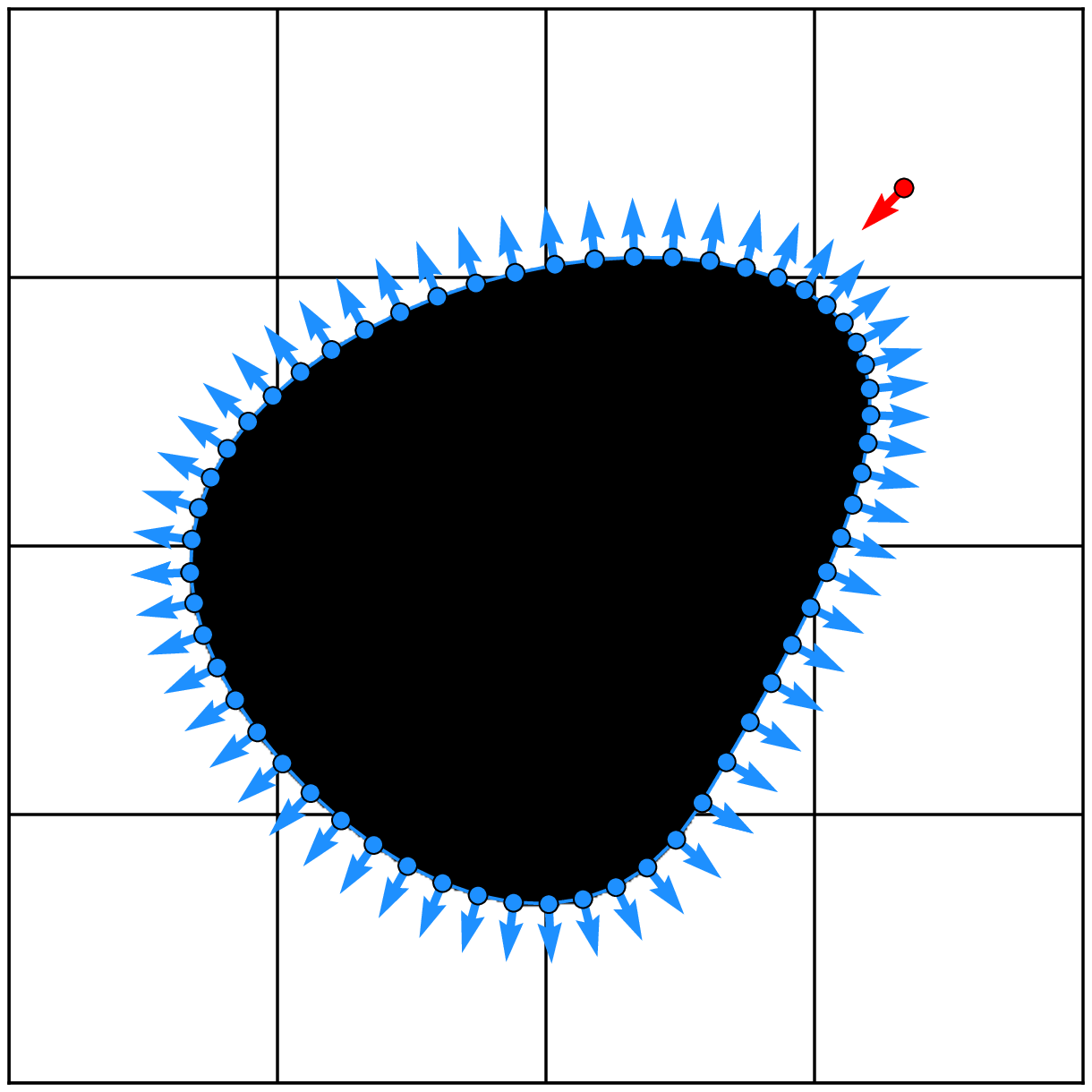} 
  }%
  \caption{\textit{The effect of a single outlier (red point) on point-membership tests. The black and white regions represent parts detected as inside or outside, respectively. }}
\end{figure}

While the cleaning process is able to remove the bulk of the outliers, if isolated outliers remain in the cloud, they may have an effect on the point-membership classification process of Algorithm~\ref{alg:ioStateInCloud}.
To demonstrate this, Figure~\ref{fig:outlier1} shows the inside-outside state recovered from a point cloud consisting of 90 points and an outlier in the upper right quadrant of the domain.
As can be seen in the figure, the presence of the outlier causes a region in the domain to be falsely detected as ,,inside''.
To deal with isolated outliers, the point membership test of Algorithm~\ref{alg:ioStateInCloud} can be modified by testing in the n-neighborhood of the query point. 
In this process, instead of checking against a single closest point, the $n$ nearest points of $\boldsymbol{q}$ are found and the point membership with respect to each of them is computed. 
If $\boldsymbol{q}$ lies inside with respect to the majority of the points in the $n$ neighborhood, its membership is determined as inside, otherwise as outside.
Following this idea, Figure~\ref{fig:outlier2} depicts the inside-outside state recovered using 3 nearest neighbor queries.
The choice of $n$ for the number of nearest neighbor queries is a parameter that needs to be determined prior to the simulation. 
In the examples of Section~\ref{ssec:3dexamples}, the parameter is chosen in the range $n=5...50$.

\section{Numerical examples}
This section demonstrates the proposed point-cloud-based FCM approach with numerical examples in two and three dimensions.
First, we study an example with a known reference solution, where all the boundary conditions are aligned with the finite cell boundaries.
This way, the modeling errors due to the approximate application of Neumann BC-s (Equation~\ref{eq:discreteNeumann}) can be ruled out.
This is followed by an example where the performance of applying Neumann boundary conditions on non-conforming interfaces (as explained in Section~\ref{ssec:pointCloudBC}) is tested.
Finally, the point cloud-based FCM method is demonstrated on actual three-dimensional examples of historical structures.
\subsection{2D studies}
\subsubsection{Perforated plate with circular hole}
\label{ssec:plateWithAHole}
\begin{figure}[htbp!]
  \centering
  \subfloat[Geometry and boundary conditions\label{fig:plateWithAHoleContinuous}.]{
    \psfragfig[width=0.5\textwidth]{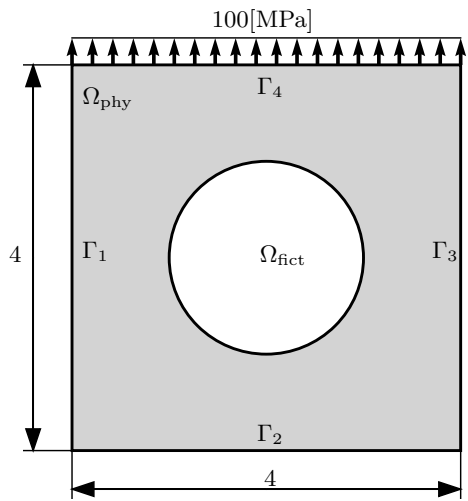} 
    {
      \footnotesize
      \psfrag{f}{$4$}
      \psfrag{g1}{$\Gamma_1$}
      \psfrag{g2}{$\Gamma_2$}
      \psfrag{g3}{$\Gamma_3$}
      \psfrag{g4}{$\Gamma_4$}
      \psfrag{h}[cc]{$100\text{[MPa]}$}
      \psfrag{op}{$\Omega_{\text{phy}}$}
      \psfrag{of}{$\Omega_{\text{fict}}$}
    }
  }%
  \subfloat[The inner boundary replaced by an oriented point cloud of 16 points.\label{fig:plateWithAHolePointCloud}]{
  \psfragfig[width=0.5\textwidth]{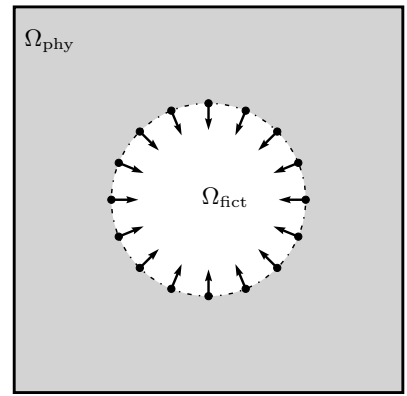} 
  {
    \footnotesize
    \psfrag{op}{$\Omega_{\text{phy}}$}
    \psfrag{of}{$\Omega_{\text{fict}}$}
  }
  }
  \caption{\textit{Rectangular plate with a circular hole}}
\end{figure}

Figure~\ref{fig:plateWithAHoleContinuous} depicts a rectangular plate with a circular hole in the center.
The plate is subjected to a constant traction along $\Gamma_4$, while symmetry boundary conditions are applied on $\Gamma_1$ and $\Gamma_2$.
The Young's modulus and the Poisson's ratio of the material are $E=2.069\cdot 10^5 \text{[MPa]}$ and $\nu = 0.29$, respectively.
Considering plane stress physics, the reference strain energy obtained by an overkill FEM analysis is $U_\text{ref}=0.7021812127$~\cite{Parvizian:2007}.
For the FCM, the embedding domain is discretized into $2 \times 2$ elements, and a value of $\alpha = 10^{-12}$ is applied to scale the material parameters in the fictitious domain.
The polynomial order of the shape functions is $p=12$.
The continuous curve representing the circular hole is replaced by an oriented point cloud consisting of $n$ points, where $n$ is gradually increased in the range of $n=4...4096$.
Refer to Figure~\ref{fig:plateWithAHolePointCloud} for an example of such a cloud.
The accuracy of the analysis is measured using the following error norm:
\begin{equation}
  e = \frac{\left|U_\text{ref}-U_\text{num}\right|}{U_\text{ref}},
\end{equation}
where $U_\text{num}$ is the strain energy computed on the discrete point cloud representation of the circular hole.

The discretization of the circular interface into an oriented point cloud can be thought of as a replacement of the boundary by an $n$-sided polygon.
This introduces a modeling error when integrating the term over $\Omega_\text{phy}$ in Equation~\ref{eq:virtualWork}, as the integration is not performed over an exact circle but rather over its polygonal approximation.
Therefore, the error is expected to converge at the same rate as the area of an $n$-sided polygon converges towards the area of its inscribed circle, i.e. quadratically.

\begin{figure}[!htbp]
  \centering
  \includegraphics[width=0.5\textwidth]{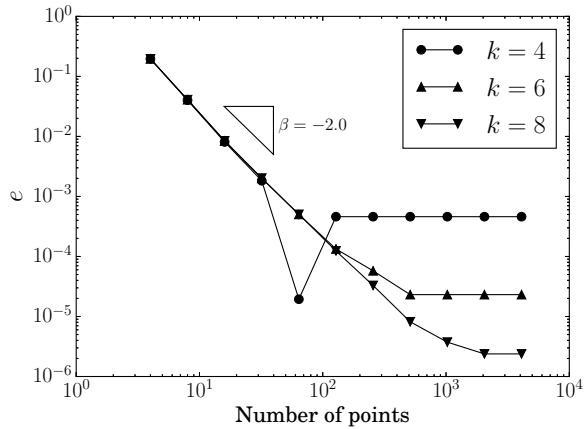}
  \caption{\textit{Rectangular plate with circular hole: convergence of error for increasing cloud densities and different maximum levels of quadtree subdivision $k$.}}
  \label{fig:plateWithAHoleConvergence}
\end{figure}

This expectation is confirmed by the error plots on Figure~\ref{fig:plateWithAHoleConvergence}.
It depicts the evolution of the error for increasing cloud densities, for different maximum levels of quadtree subdivision~${k=\left\{4,6,8\right\}}$, see Section~\ref{ssec:fcm}. 
Initially, the polygonal approximation dominates the error, and the curves show quadratic convergence.
Eventually, depending on the value of $k$, the convergence curves level off into a plateau.
In the plateau region, the error due to the quadtree-based integration overtakes the polygonal discretization error: even though the interface is modeled by higher resolutions, the integration scheme is not able to resolve this increase in geometric accuracy.

\subsubsection{Perforated plate with elliptical hole under internal pressure}
\begin{figure}[!htbp]
  \centering
  \psfragfig[width=0.6\textwidth]{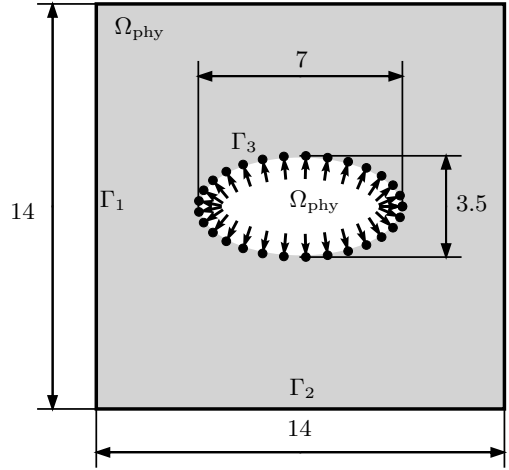} 
  {
    \footnotesize
    \psfrag{l}[cc]{$14$}
    \psfrag{a}{$7$}
    \psfrag{b}{$3.5$}
    \psfrag{g1}{$\Gamma_1$}
    \psfrag{g2}{$\Gamma_2$}
    \psfrag{g3}{$\Gamma_3$}
    \psfrag{op}{$\Omega_{\text{phy}}$}
    \psfrag{of}{$\Omega_{\text{fict}}$}
  }
  \caption{\textit{Rectangular plate with elliptical hole under internal pressure.}}
  \label{fig:plateWithEllipticalHole}
\end{figure}

In this example, the circular interface from the previous section is replaced by an elliptical curve~(Figure~\ref{fig:plateWithEllipticalHole}).
To investigate the effects of applying Neumann boundary conditions as described in Section~\ref{ssec:pointCloudBC}, the elliptical hole is discretized into an oriented point cloud along which a constant 
internal pressure of $1\text{[MPa]}$ is applied. 
The reference value for the strain energy computed by p-FEM is $U=44.28375067893$.
The domain is discretized into a regular grid of $6\times6$ finite cells, where the polynomial order of the shape functions varies in the range $p=1...10$.
The regularization parameter of the Dirac delta function in Equation~\ref{eq:regularizedDiracDelta} is chosen to be $\epsilon=0.0625$.
The diffuse region obtained in this way (see Figure~\ref{fig:pcaDistance}) is integrated using a composed Gaussian quadrature of order $10$ combined with a quadtree-based subdivision of maximum level $k=8$.
While this integration depth seems to be excessively large for practical applications, it should be noted that the size of the surfaces for which weak boundary conditions need to be applied is usually significantly smaller than the overall sizes of the geometries of interest. 
Therefore, applying the boundary conditions in this manner does not lead to a significant performance penalty.

To rule out the errors associated to the discontinuity in the indicator function $\alpha(\boldsymbol{x})$, the stiffness matrix is integrated using the highly precise integration technique described in~\cite{Kudela:2015}. 
This way, the error due to the approximate application of the Neumann boundary condition is examined exclusively.

\begin{figure}[h!]
  \centering
  \includegraphics[width=0.6\textwidth]{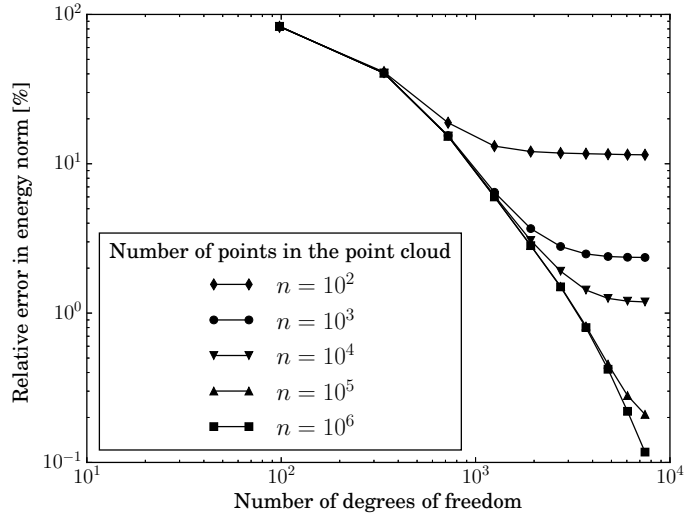}
  \caption{\textit{Convergence of the error in energy norm when boundary conditions are applied using the regularized delta function. The polynomial order of shape functions is varied in the range $p=1..10$. The regularization parameter of the delta function is $\epsilon=0.0625$. }}
  \label{fig:boundaryConditionConvergence}
\end{figure}

Figure~\ref{fig:boundaryConditionConvergence} depicts the convergence of the error in the energy norm for different point cloud densities.
As the figure shows, the expected exponential rate of convergence can be attained in the pre-asymptotic range.
However, similar to the study conducted in Section~\ref{ssec:plateWithAHole}, the convergence curves level off into a plateau, depending on the density of the point cloud.
As higher cloud densities are able to represent the underlying elliptical contour with higher accuracy, the level-off location shifts towards lower errors for an increasing number of points in the cloud.
It is noted here, however, that even a relatively low density (1000 points) is able to produce an error in the range of $1\%$, which is sufficient for most engineering applications.

\subsection{3D examples}
\label{ssec:3dexamples}
In the following, the proposed point-cloud-based FCM approach is demonstrated on three-dimensional structures represented by oriented point clouds.
\subsubsection{Athlete}
\begin{figure}
  \centering
  \subfloat[Example input picture\label{fig:statue}]{
    \includegraphics[width=0.2\textwidth]{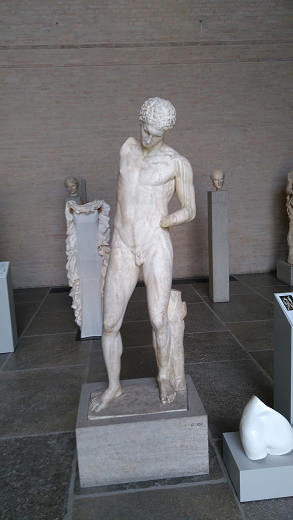}
  }
  \subfloat[Recovered point cloud and camera positions\label{fig:cloudAndCameras}]{
    \includegraphics[width=0.5\textwidth]{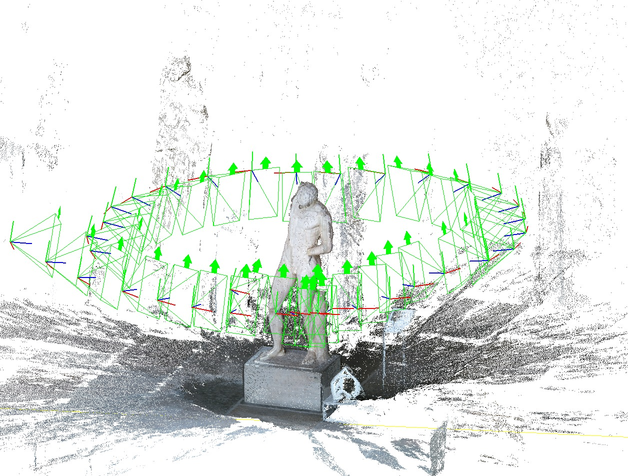}
  }
  \subfloat[Point cloud and the associated normal vectors\label{fig:cloudAndNormals}]{
    \includegraphics[width=0.2\textwidth]{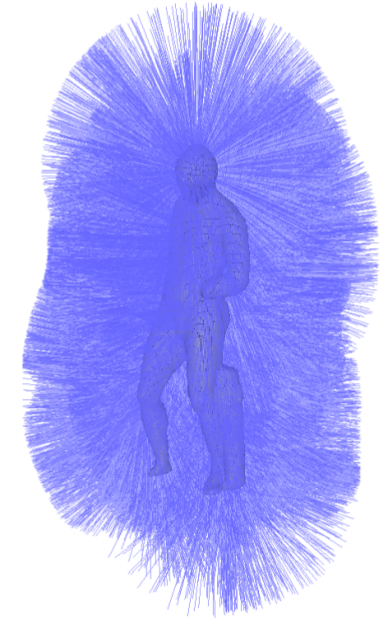}
  }
  \caption{\textit{Statue example: input pictures and the resulting cloud}}
\end{figure}

\begin{figure}[!htbp]
  \centering
  \subfloat[Point cloud embedded into a regular mesh of finite cells\label{fig:statueEmbedded}]{
    \qquad
    \includegraphics[width=0.2\textwidth]{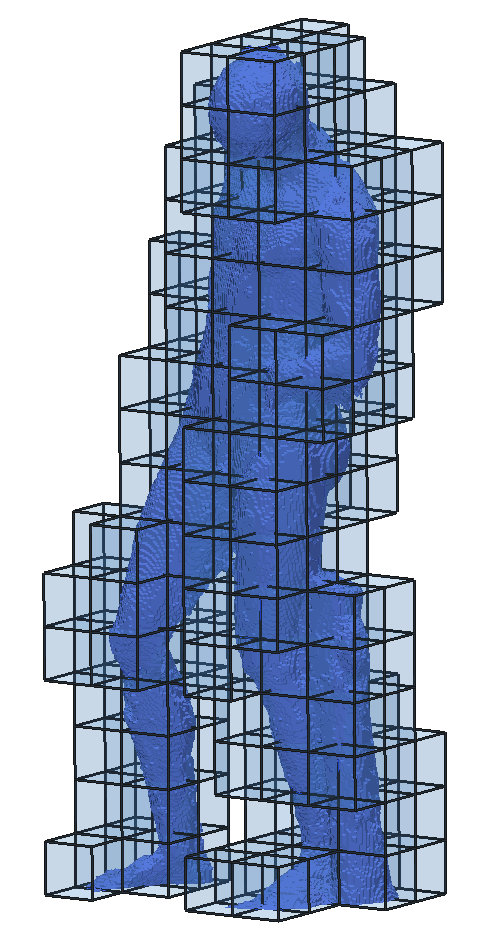}
    \qquad
  }
  \qquad
  \subfloat[von-Mises stresses throughout the structure\label{fig:statueStressField}]{
    \qquad
    \includegraphics[width=0.17\textwidth]{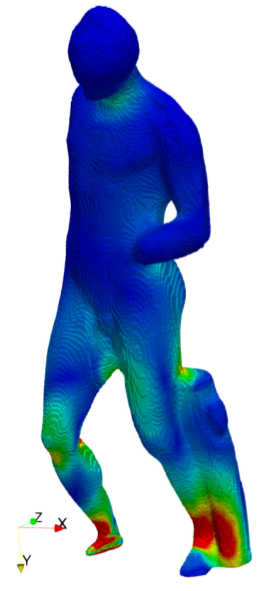}
    \qquad
  }
  \qquad
  \subfloat[A detailed view on a section of the left foot\label{fig:statueStressFieldFoot}]{
    \includegraphics[width=0.2\textwidth]{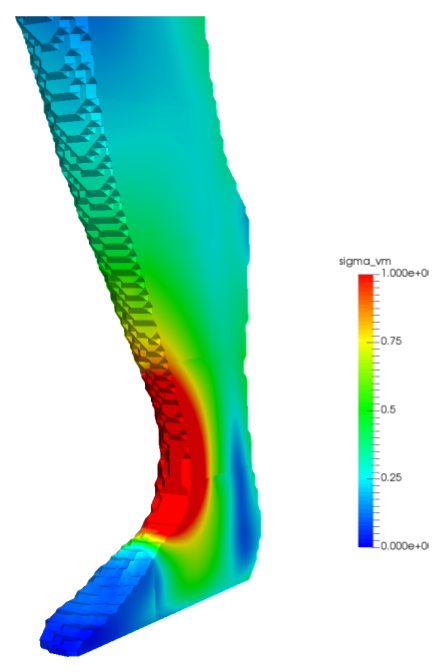}
  }
  \caption{\textit{Statue example: discretization and stresses}}
\end{figure}

Figure~\ref{fig:statue} shows a statue from the museum ,,Glyptothek'' located in Munich, Germany.
Images of the object were taken from 36 different angles, using a cell phone camera.
These input images were processed using the popular structure-from-motion toolbox VisualSFM~\cite{Wu:2011,Wu:2013} and the multi-view reconstruction algorithm of~\cite{Furukawa:2010}, as demonstrated in Figure~\ref{fig:cloudAndCameras}.
The resulting point cloud is depicted in Figure~\ref{fig:cloudAndNormals}.
The point cloud was embedded in a regular mesh of $325$ finite cells with polynomial order $p=5$, as shown in Figure~\ref{fig:statueEmbedded}.
Linear elastic material behavior is assumed, and the structure is loaded under its self-weight.
The scaling factor $\alpha$ for the FCM was chosen as $10^{-6}$.
Homogenous Dirichlet boundary conditions were applied on the bottom faces of the finite cell mesh, in order to rigidly fix the statue to the ground.

The resulting stress field is depicted in Figure~\ref{fig:statueStressField}, while Figure~\ref{fig:statueStressFieldFoot} shows a detailed view of a cross-section of the left foot. 
As seen in the Figure, the peak stress occurs at the ,,ankle'' region, which possesses the smallest cross-section over the entire structure.
This phenomenon is in good accordance with other observations from the study of the structural behavior of stone statues: numerical computations conducted on ,,David'' from Michelangelo showed a similar stress distribution, with the peak occurring in the ankle area~\cite{Borri:2006}.
Interestingly, there are also other areas of increased stresses -- such as the area around the neck as well as areas in the upper thigh. 
Other intuitive candidates for high stresses, such as the left arms and shoulders can be disregarded. 

While this example illustrates the main steps of the proposed point-cloud-based FCM pipeline, without further knowledge about the material parameters, the computed results merely allow for a qualitative assessment of the stress distribution in the statue, under the assumption that no internal cavities are present and that the material is uniform.
The next example addresses this question on a structure fow which the material properties are known.
\subsubsection{The cistern of the Hagia Thekla Basilica in Turkey}
\label{ssec:cistern}
The archaeological site at Hagia Thekla (Meryemlik, today Turkey) was a major pilgrimage site in late antiquity~\cite{Hill:1996}.
The site features numerous above-ground structures. 

\begin{figure}[!htbp]
  \centering
  \subfloat[Interior view.\label{fig:hagiaTeklaPhoto}]{
  \includegraphics[width=0.4\textwidth]{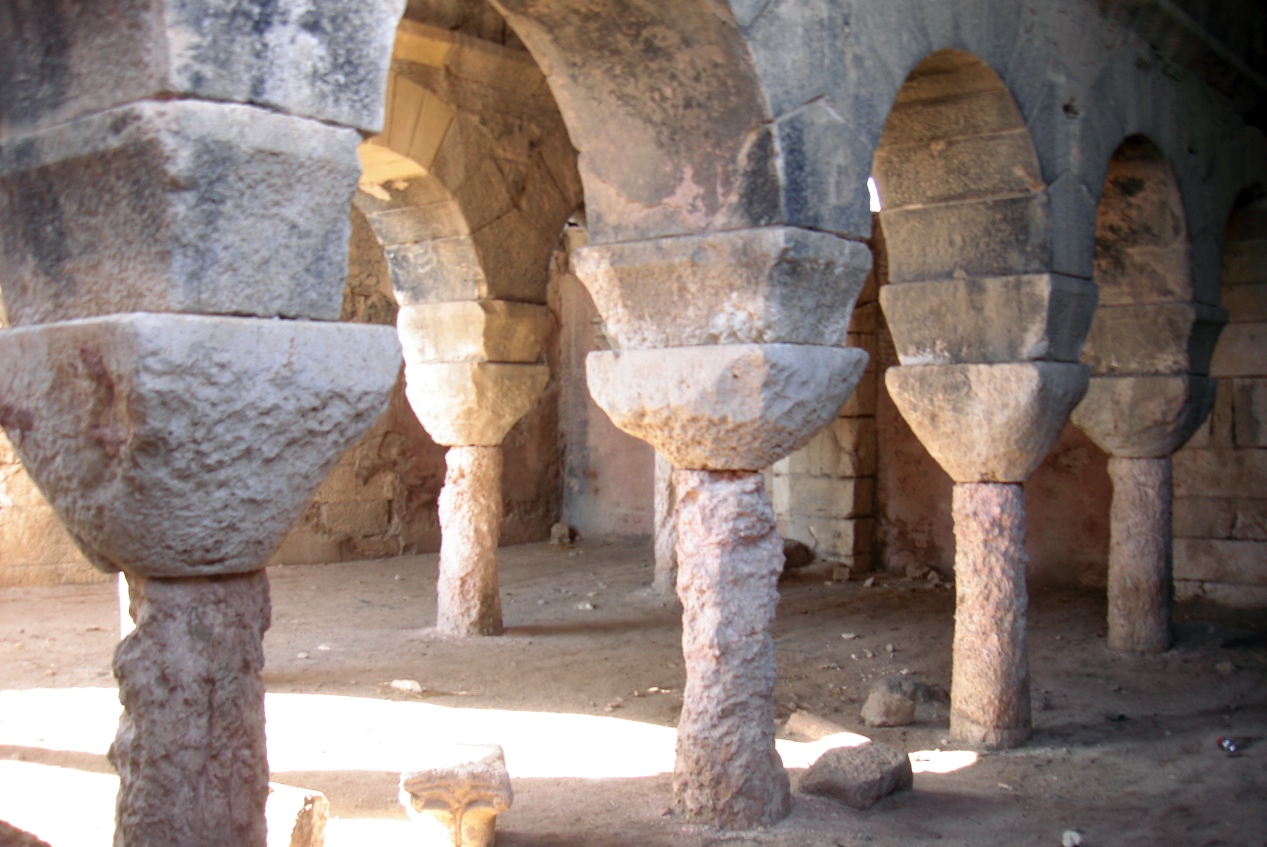}
  }
  \subfloat[Plan and cross section.\label{fig:hagiaTeklaPlan}]{
  \includegraphics[width=0.6\textwidth]{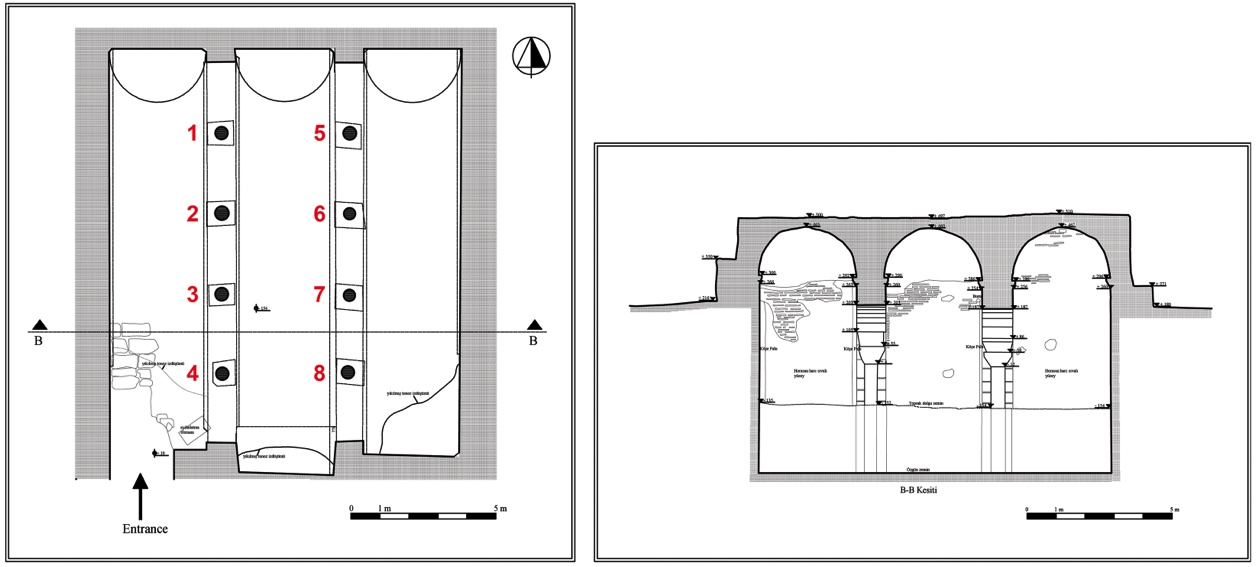}
  }
  \caption{\textit{The cistern of Hagia Thekla Basilica.}}
\end{figure}

The cistern of the Thekla Basilica is part of the water storage and distribution system of the main church of the site and its sacred surrounding area, which is enclosed by walls. 
It has a rectangular plan measuring approximately $12\times 14.6$ meters in the interior. 
The interior space is divided into three aisles by two rows of columns (Figures~\ref{fig:hagiaTeklaPlan}~and~\ref{fig:hagiaTeklaPhoto}).
The columns in each row are connected by arches. 
Three barrel vaults cover the interior running in the north-south direction. 
The columns supporting the upper structure are made of a pink calcareous stone and originally had a diameter of approximately $45$ cm.
The columns have double capitals made of limestone. 
It is not possible to make observations about the condition of the column bases and the floor, due to the thick layer of earth accumulated inside the cistern over centuries.   
The outer walls were built with a multi-leaf masonry construction system. 
The outer facing of the walls are made of big limestone blocks, while the inner faces consist of brick and mortar. 
As seen in Figure~\ref{fig:hagiaTeklaPhoto}, the cross-sections of the columns have decreased remarkably.
The exterior surfaces are flaking due to physicochemical effects; the erosion continues. 
In addition to surface erosion with a non-uniform pattern, there are deep cavities on the surfaces of the columns. 
One of the columns (Column 3) has already collapsed and was replaced by a concrete column in the 1960's.

\begin{figure}[!htbp]
  \centering
  \subfloat[Point cloud\label{fig:ayateklaCloud}]{
  \includegraphics[width=0.7\textwidth]{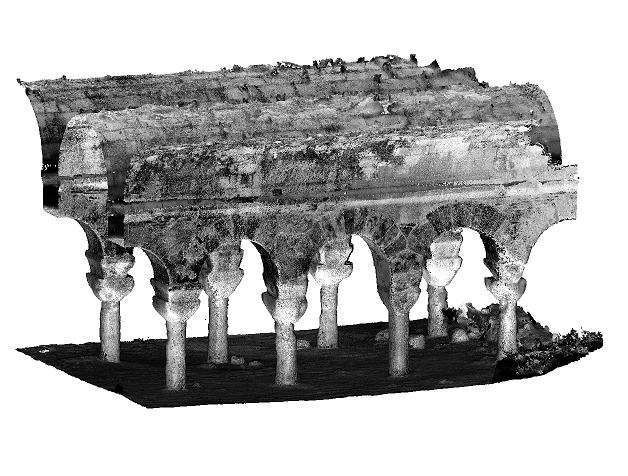}
  }
  \\
  \subfloat[The geometry embedded into a finite cell mesh\label{fig:ayateklaMesh}]{
  \includegraphics[width=0.8\textwidth]{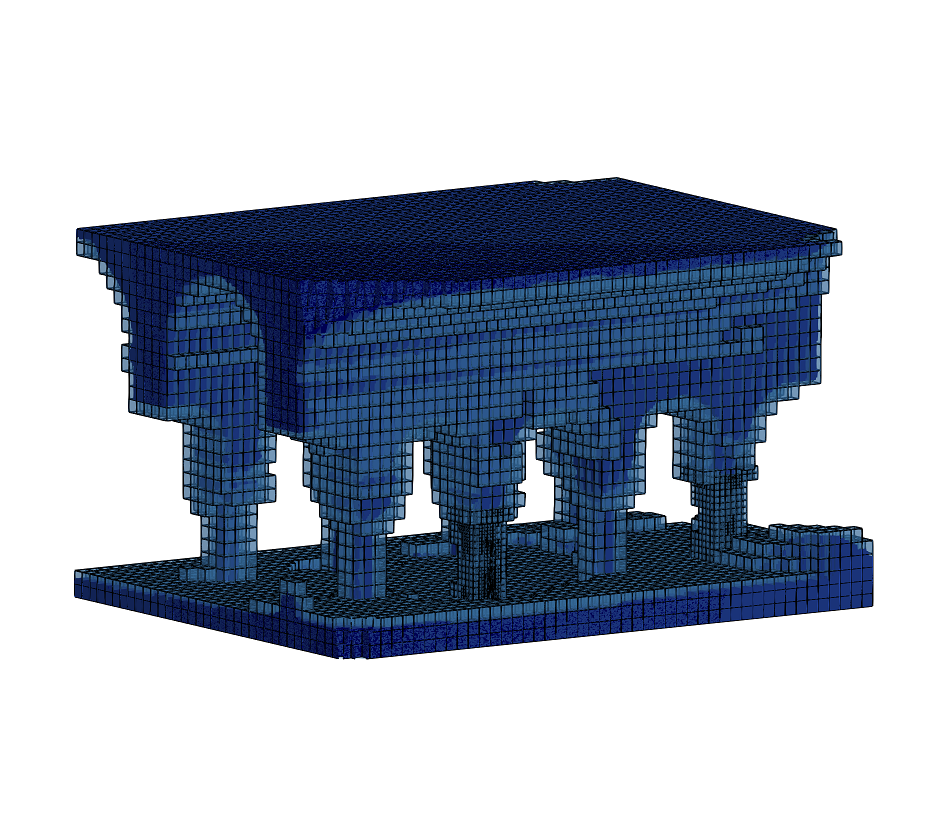}
  }
  \caption{\textit{Cistern example}}
\end{figure}

It is difficult to record the shape of the decayed column surfaces and cavities relying on manual measurement procedures.
Therefore, a high definition surveying scanner was employed to document these elements. 
During the field campaign, the instrument was set up at a number of positions around each column at a distance of a few meters. 
Thus, a maximum point density of approx. $5$ mm was ensured to represent the highly decayed columns.
More details on the measurement process can be found in~\cite{Almac:2016}.
Figure~\ref{fig:ayateklaCloud} depicts the measured point cloud, consisting of $10^7$ points.

\begin{figure}[!t]
  \centering
  \subfloat[Column 2]{
  \includegraphics[width=0.5\textwidth]{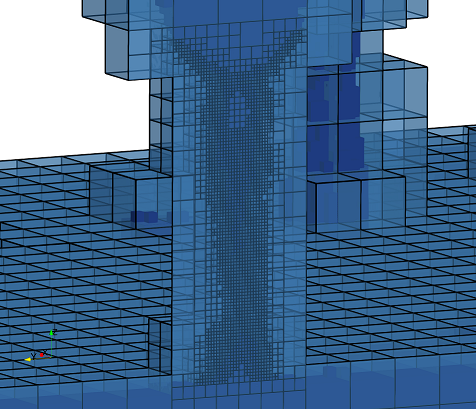}
  }
  \subfloat[Column 4]{
  \includegraphics[width=0.5\textwidth]{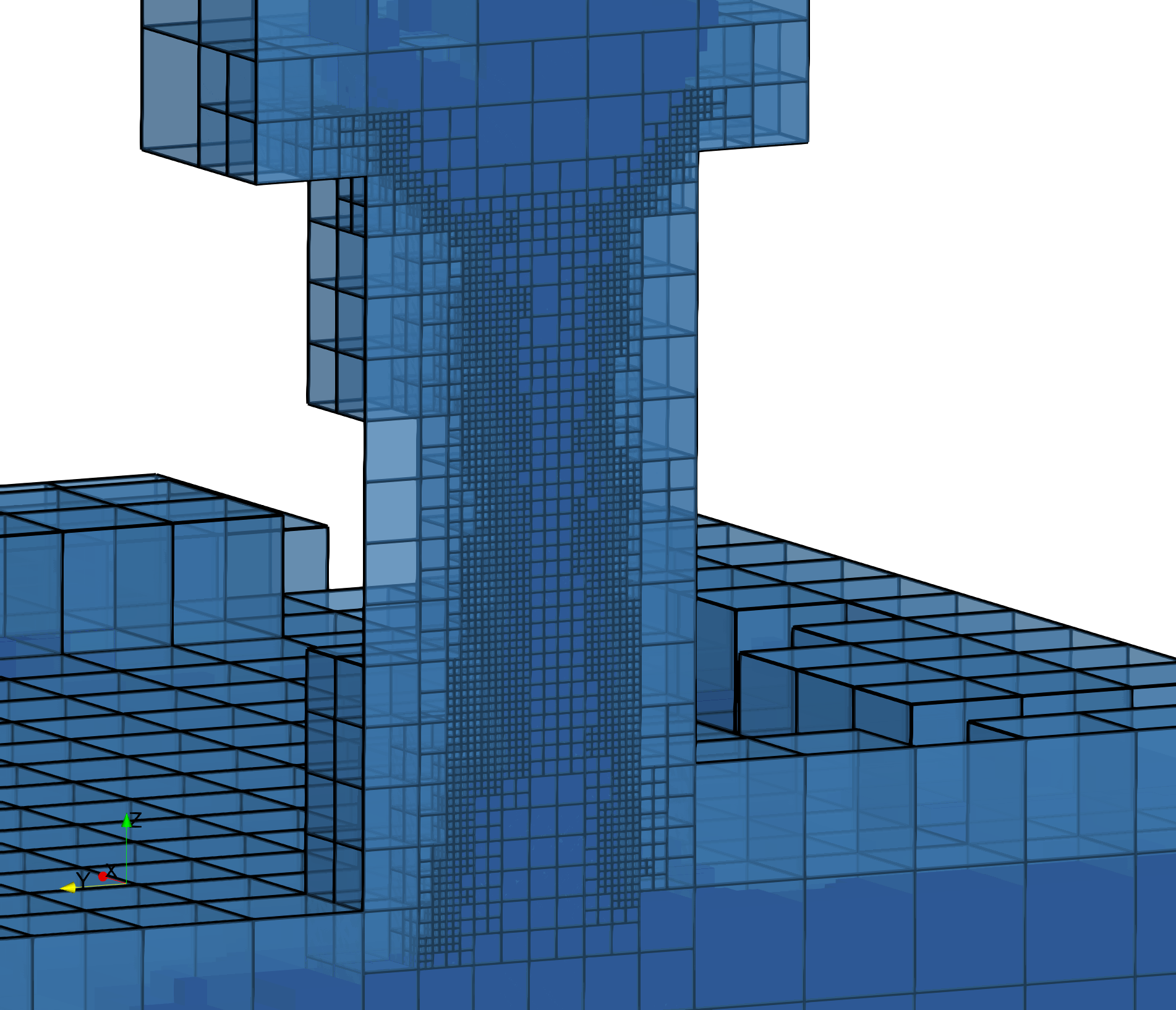}
  }
  \caption{\textit{Refined computational grid around the two columns.\label{fig:ayateklaRefinedMesh}}}
\end{figure}

The most vulnerable elements of the structure are the columns.
As stress concentrations are expected at the cavities on the surfaces of the columns, a reduction of the discretization error by a refinement of the computational grid is needed.
For reasons of efficiency, it is important to refine the grid only around the columns, where the stress field is expected to change rapidly.
For the FEM and the FCM, such \textit{local refinement} techniques have been well-studied recently. 
In our applications, we employ the \textit{multi-level $hp$-adaptivity} technique of~\cite{Zander:2015}.
In the refinement procedure, those cells that are intersected by the points representing column 2 and 4 are recursively subdivided into eight equal subcells, until a subdivision depth of 4 is reached.
A cross-sectional view of the refined mesh is depicted in Figure~\ref{fig:ayateklaRefinedMesh}.

\begin{figure}[!htbp]
  \centering
  \subfloat[Complete structure]{
  \includegraphics[width=0.7\textwidth]{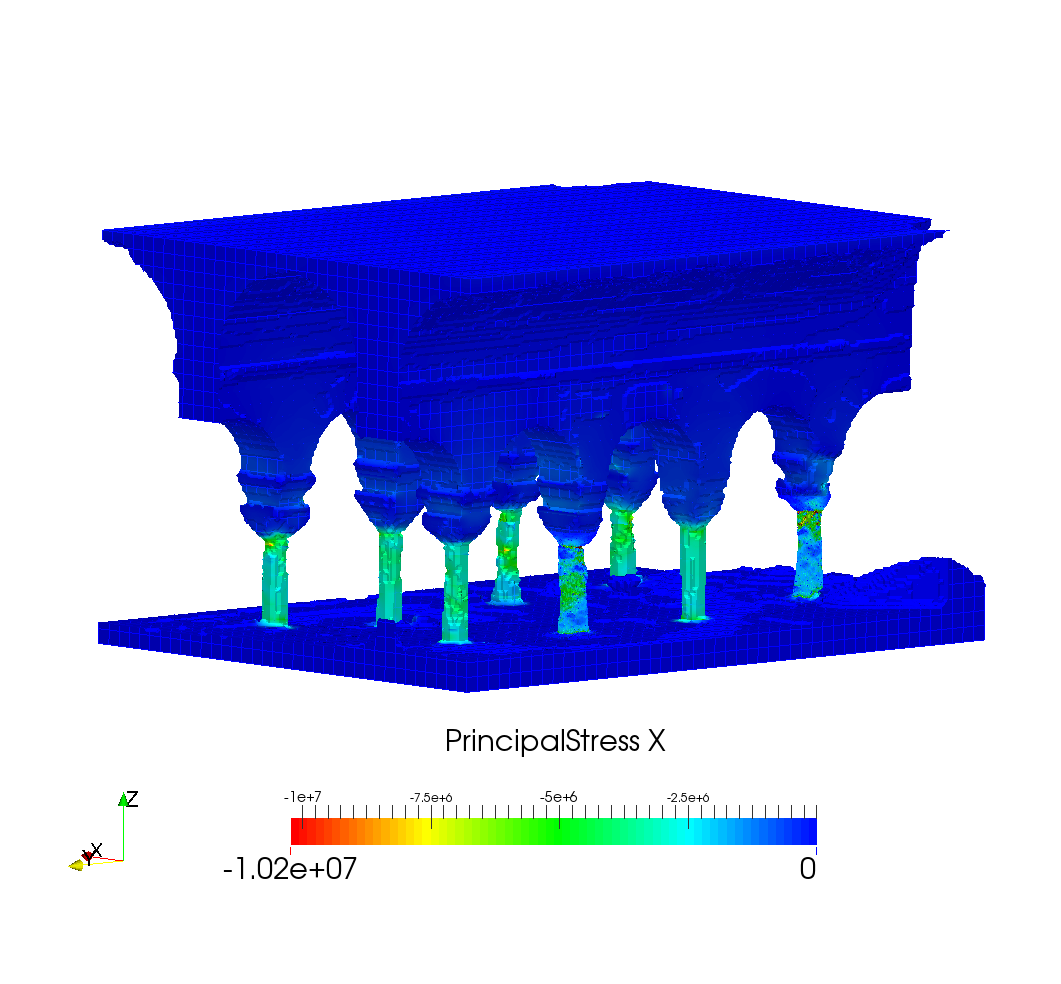}
  }
  \\
  \subfloat[Principal stresses in Column 2]{
  \includegraphics[width=0.5\textwidth]{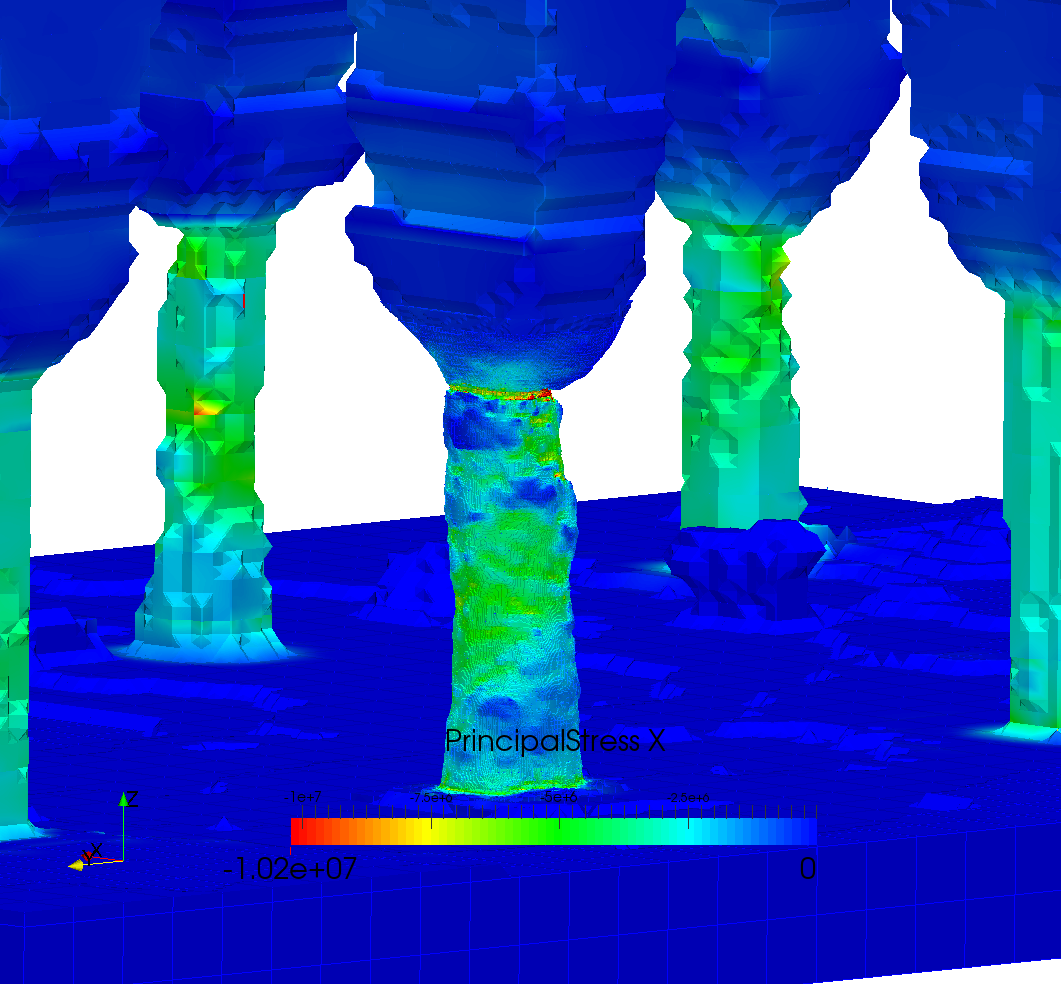}
  }
  \subfloat[Principal stresses in Column 4]{
  \includegraphics[width=0.5\textwidth]{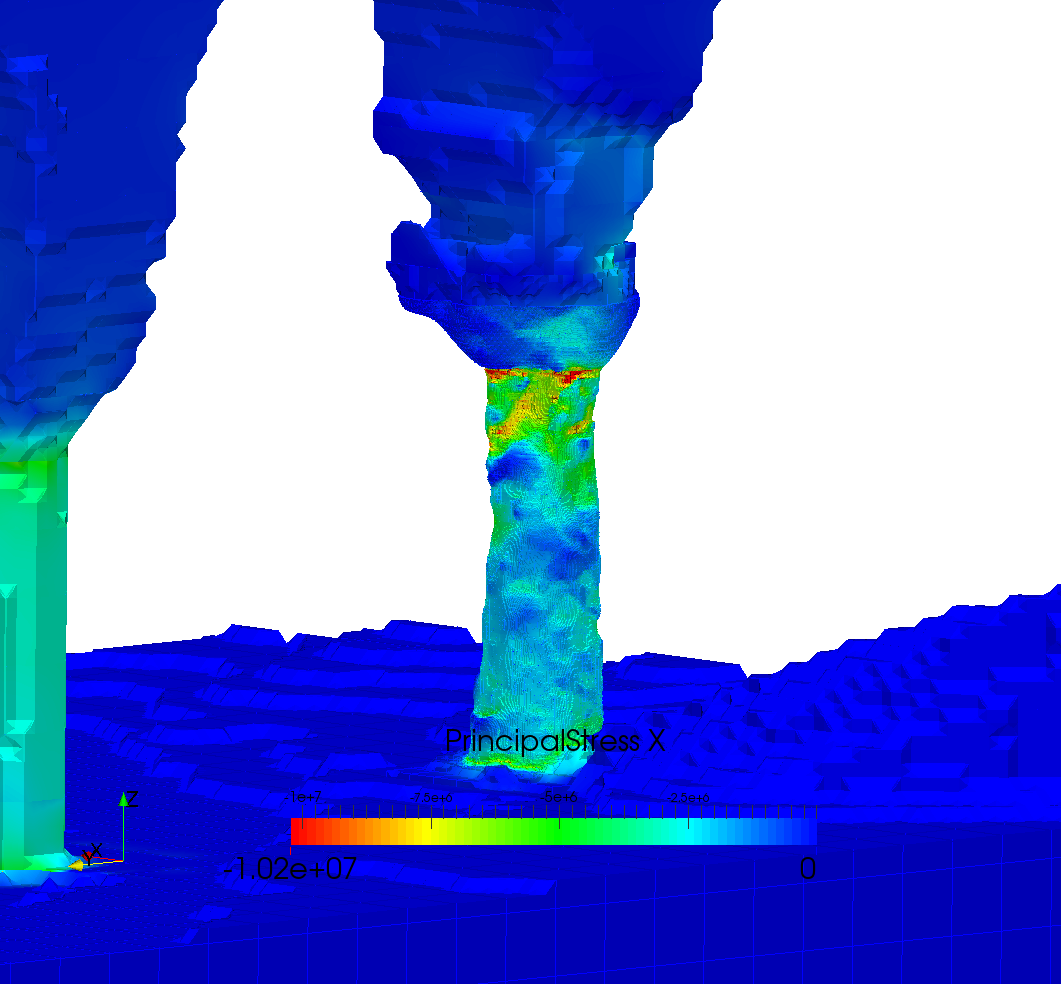}
  }
  \caption{\textit{Cistern example: principal stress distribution}}
  \label{fig:princStress}
\end{figure}

The material properties were defined to be linear elastic and isotropic, with an elastic modulus and a Poisson's ratio of $E=2\cdot 10^4 \text{MPa}$ and $\nu=0.2$, respectively.
The specific gravity of the material was set to $27\text{ kN}/m^3$.
In the fictitious domain, the material was given a stiffness of $2\cdot 10^{-4} \text{ MPa}$.
The foundation of the structure was rigidly fixed to the ground.
The maximum principal stress distribution computed by the FCM is depicted in Figure~\ref{fig:princStress}.

\begin{figure}[!htbp]
  \centering
  \includegraphics[width=0.9\textwidth]{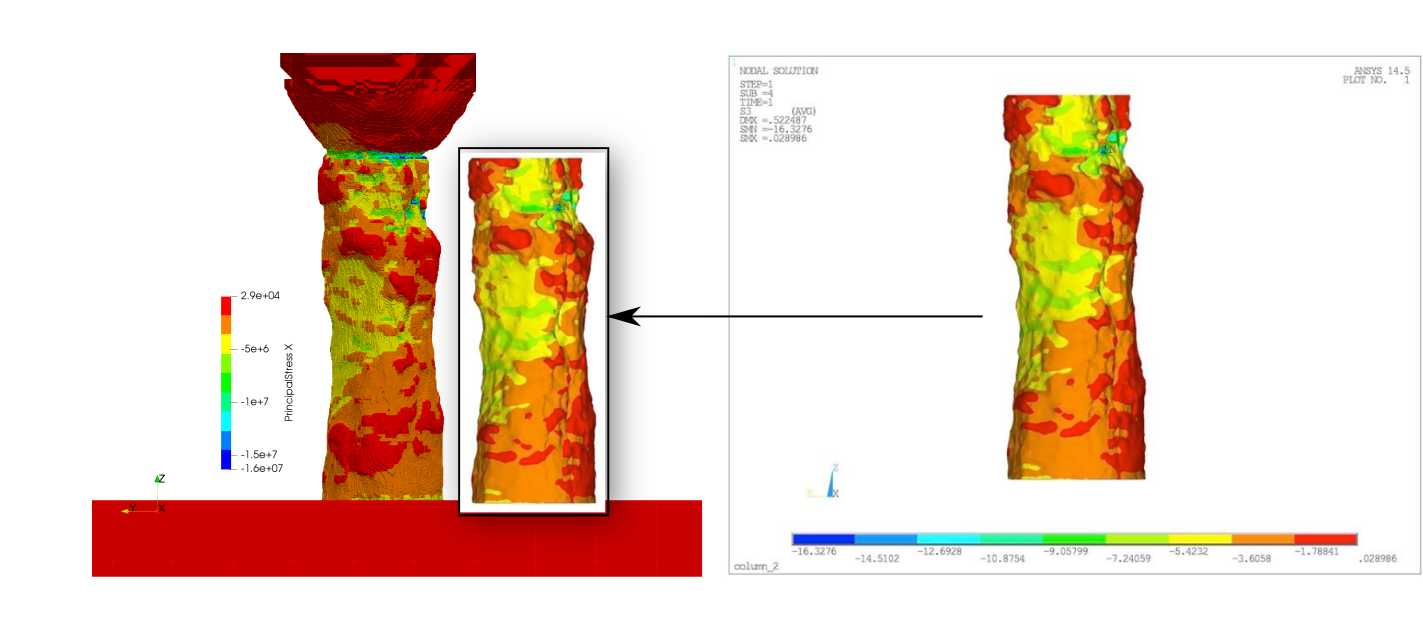}
  \caption{\textit{Cistern example: comparison of the maximum principal stresses computed by the FCM (left) and a commercial FEM software (right). The picture on the right is taken from~\cite{Almac:2016}.}}
  \label{fig:compareToAnsys}
\end{figure}

As expected, the highest compressive stresses occur in the columns.
The stress values are in the range of $2...6 \text{ MPa}$, while the peak value occurs at the connection between the columns and their capitals.
This is in good agreement with the values computed in \cite{Almac:2016}, following the traditional measurement-to-analysis procedure, i.e. reconstructing the surface, meshing the volume and performing a finite element analysis.
A comparison of principal stresses along column 2 is given in~Figure~\ref{fig:compareToAnsys}.
As seen in the Figure, the stress patterns show high resemblance. 
However, the results computed by our technique neither required the recovery of a geometric surface model, nor the generation of a boundary-conforming finite element mesh, allowing for performing the analysis directly on the point cloud. 

\begin{figure}
  \centering
  \includegraphics[width=0.7\textwidth]{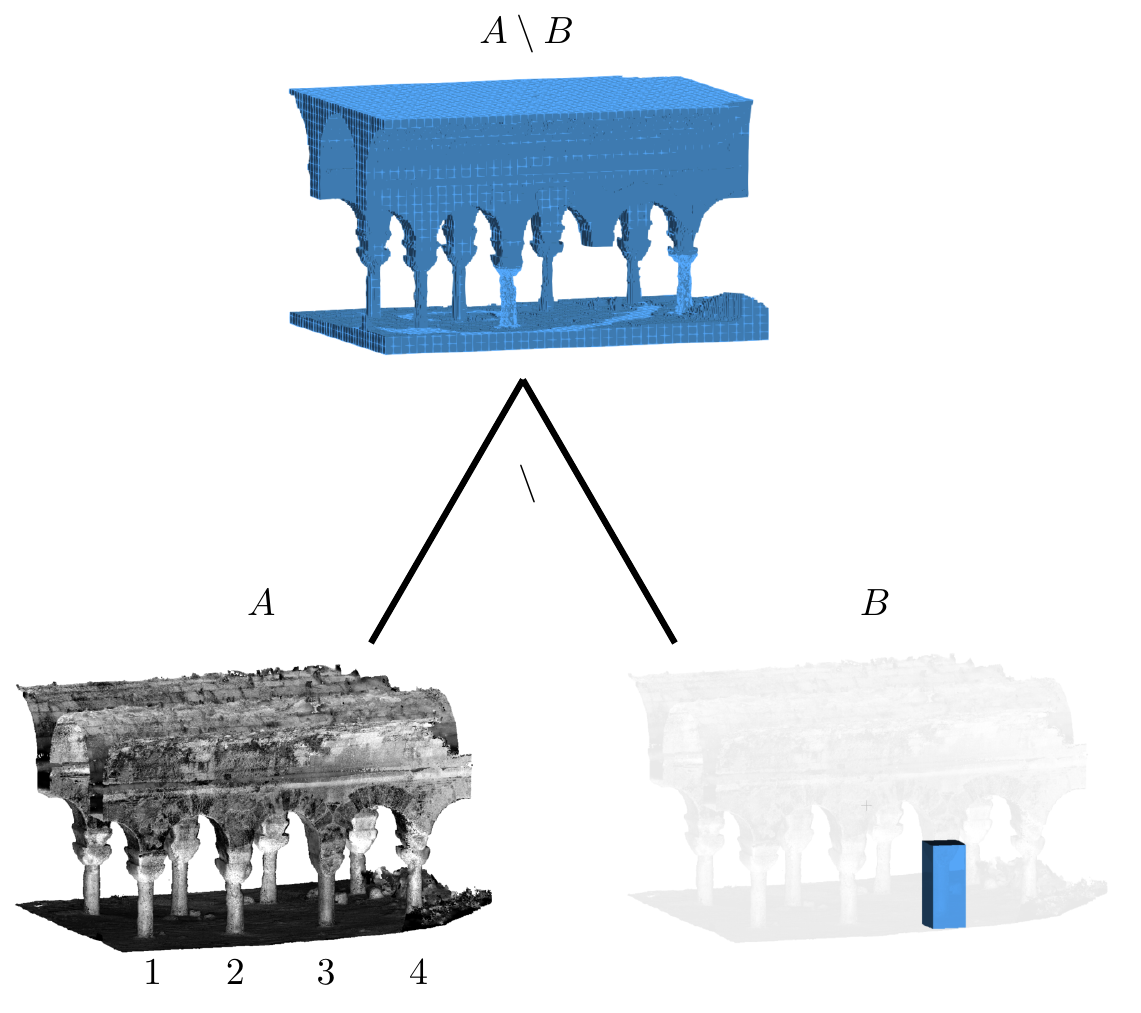}
  \caption{\textit{Column 3 removed using a CSG tree with a boolean difference operation. The column numbers are displayed under the point cloud.}}
  \label{fig:missingColumn}
\end{figure}

As the only requirement toward the geometric model is that it needs to be able to provide point-membership information, the finite cell method opens up a convenient way to investigate the effects of geometric changes, such as the removal of a specific column.
Within the FCM, it is easily possible to integrate different geometric models following the idea of Constructive Solid Geometry (CSG) modeling~\cite{Wassermann:2017}.
In CSG, 3D objects can be created by combining geometric primitives into a tree structure, using boolean operations.
To determine if a point lies within the model or not, the tree structure is traversed from the root towards the leaves, combining the inside-outside state according to the boolean operations at each level.
Following this idea, we investigate the effect of removing Column 3 from $\Omega_{\text{phy}}$: using boolean difference, the bounding box of the points representing the column is subtracted from the point cloud, as shown in Figure~\ref{fig:missingColumn}.

\begin{figure}[!htbp]
  \centering
  \subfloat[Principal stresses in Column 2]{
  \includegraphics[width=0.5\textwidth]{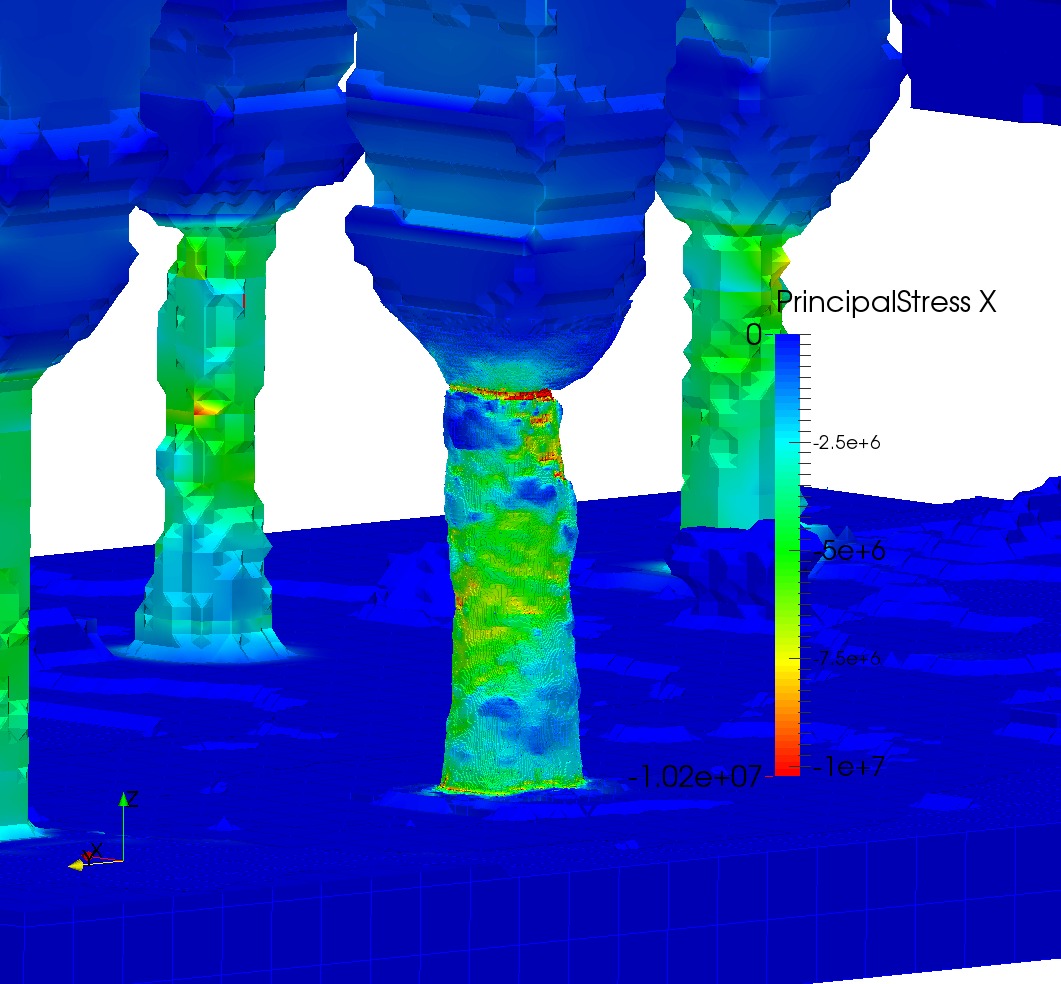}
  }
  \subfloat[Principal stresses in Column 4]{
  \includegraphics[width=0.5\textwidth]{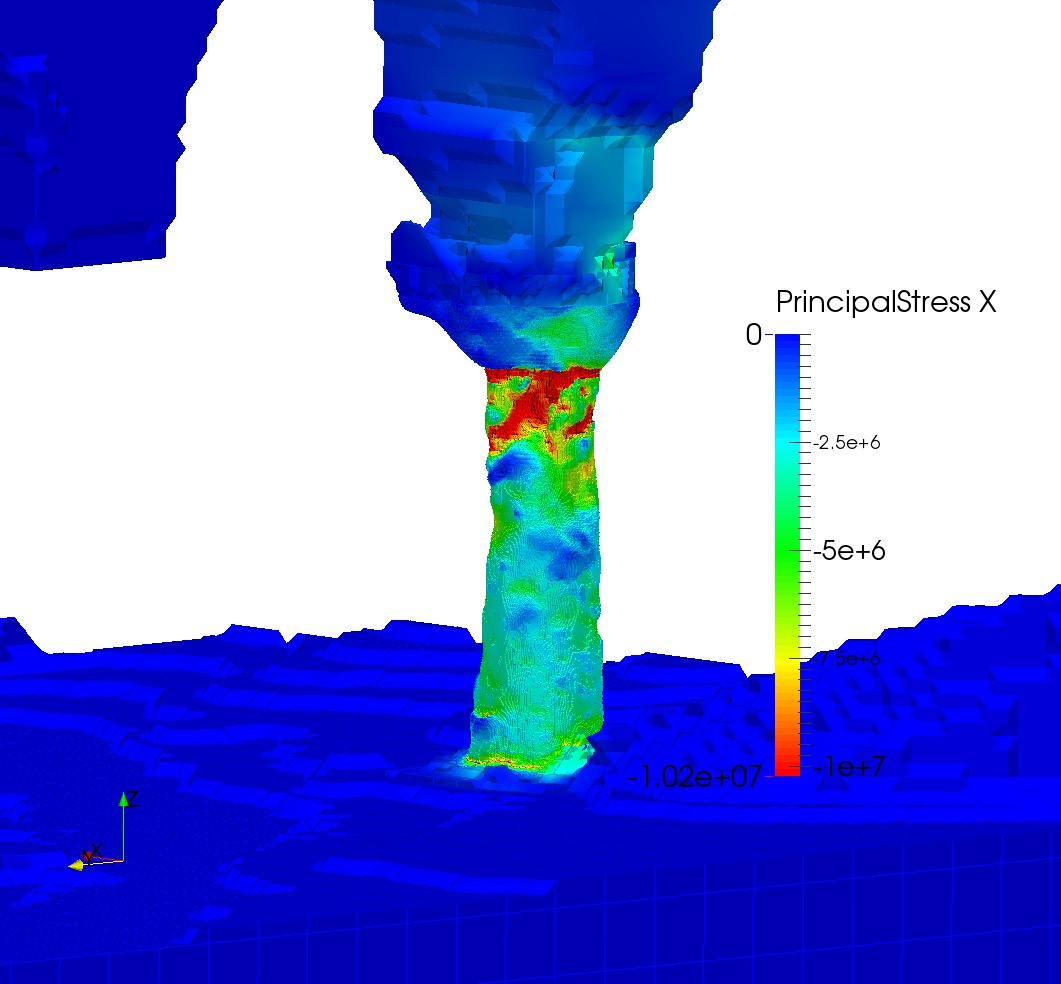}
  }
  \caption{\textit{Cistern example: principal stress distribution without Column 3}}
  \label{fig:princStressMissingColumn}
\end{figure}

\begin{figure}
  \centering
  \subfloat[Intact structure]{
    \begin{tabular}[b]{c}
    \includegraphics[width=0.5\textwidth]{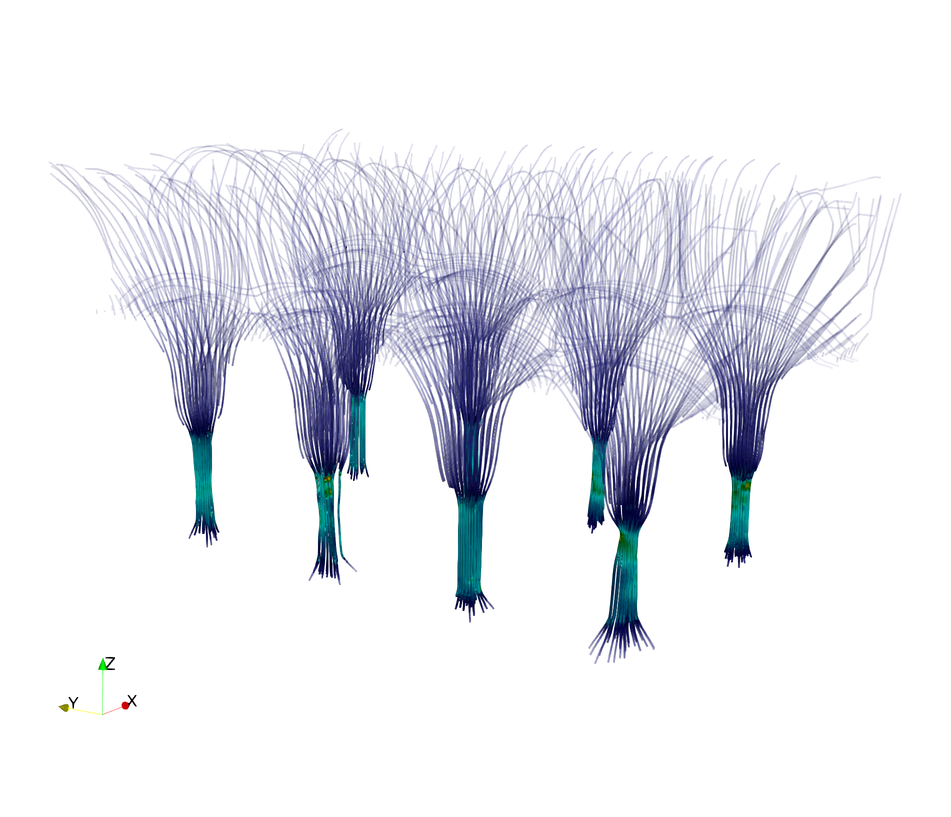} \\
    \includegraphics[width=0.5\textwidth]{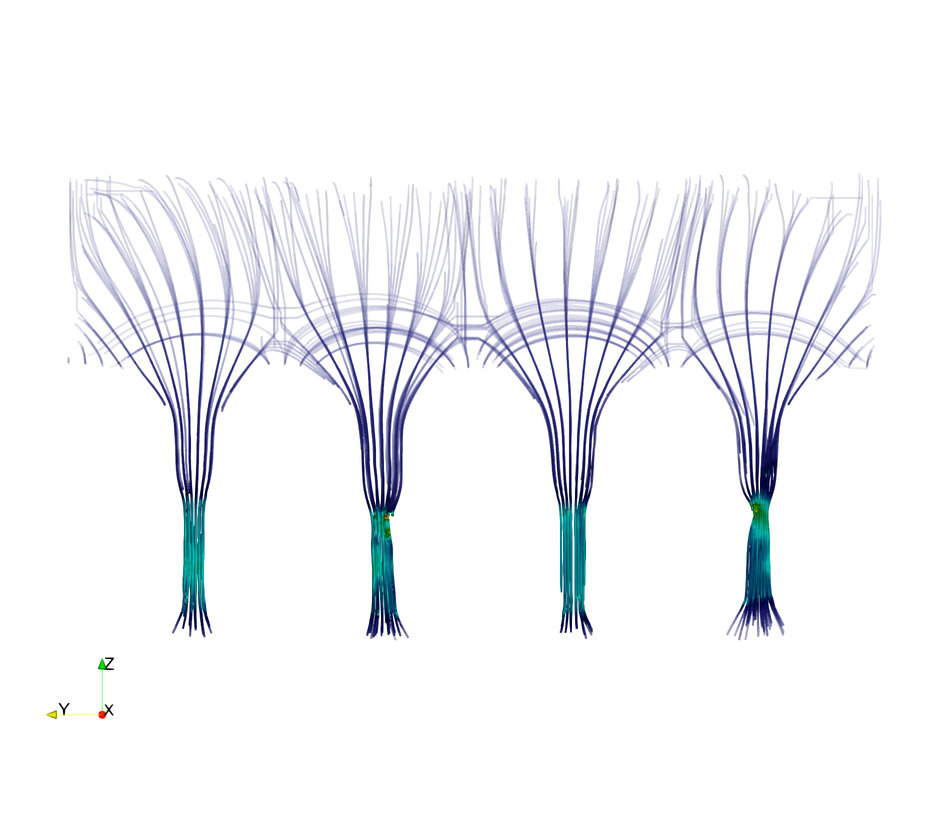}
    \end{tabular}}
  \subfloat[Structure without Column 3]{
    \begin{tabular}[b]{c}
    \includegraphics[width=0.5\textwidth]{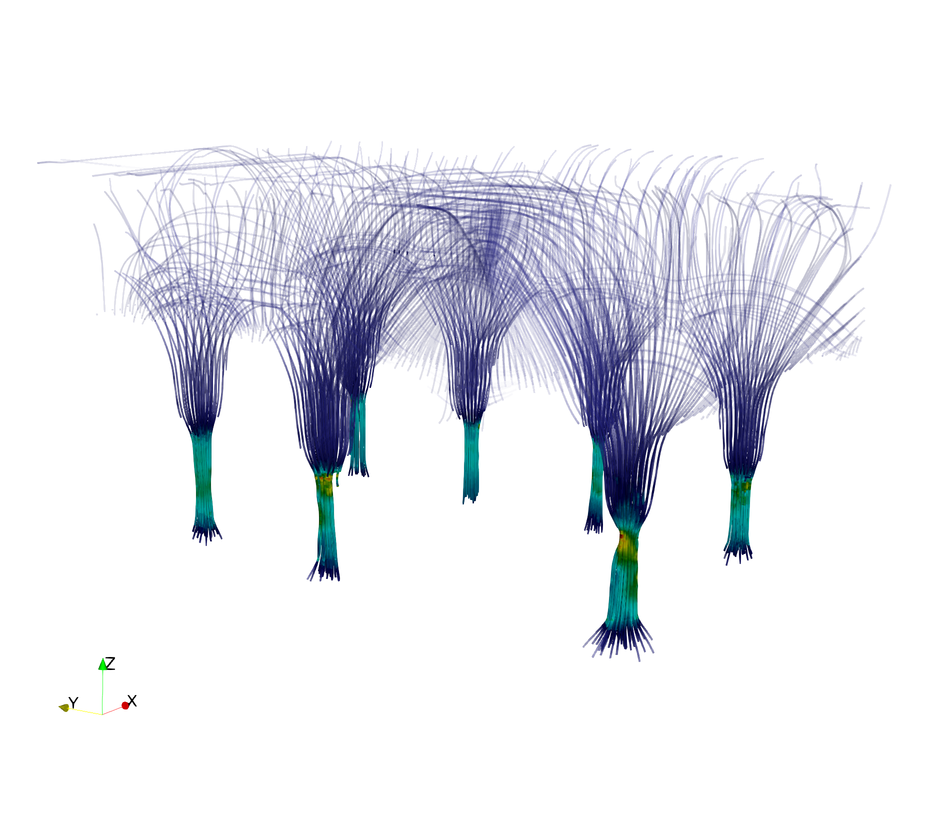} \\
    \includegraphics[width=0.5\textwidth]{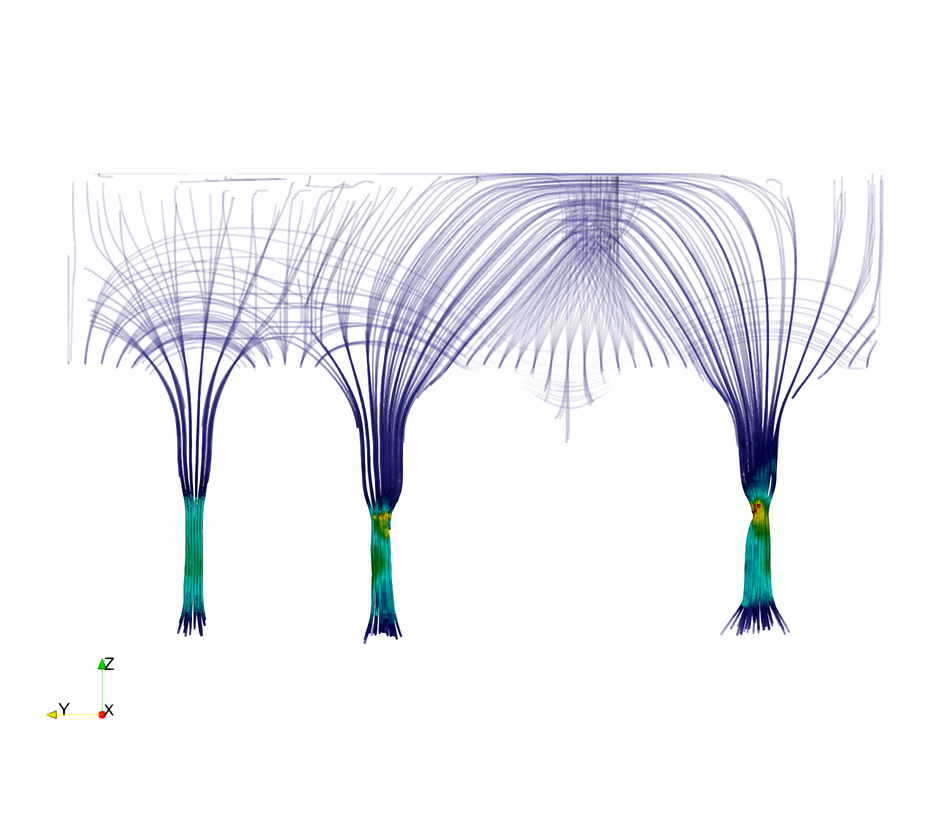}
    \end{tabular}}
  \caption{\textit{Cistern example: principal stress trajectories}}
  \label{fig:principalStressTrajectories}
\end{figure}

The principal stress distribution in this scenario is depicted in Figure~\ref{fig:princStressMissingColumn}.
As shown on the figure, the removal of the column causes a redistribution of the loads onto the neighboring supports, leading to an increase in the principal stresses in columns 2 and 4.
The redistribution phenomenon can be demonstrated by comparing the principal stress trajectories in the configurations with and without the column, as depicted in Figure~\ref{fig:principalStressTrajectories}.
Due to the removal of the column, a new ,,arc'' forms between the two neighboring columns. 
This arc is in compression and carries the redistributed load.
The newly formed stress state tells us why no structural failure occurred when column 3 collapsed: the structure was able to hold becuase stone is able to carry substantially higher loads in compression than in tension, and because the arc of principal stresses is predominantly in compression.

\section{Conclusions and outlook}
This contribution presented a method aiming at the numerical analysis of objects represented by oriented point clouds.
Instead of relying on boundary-conforming computational meshes, the approach uses the finite cell method, which computes on a rectangular background grid and only requires inside-outside information from the geometric model.
It was shown that oriented point clouds provide sufficient information to perform these point membership tests.
This allows to avoid the difficult steps of geometry recovery and mesh generation in the usual measurement-to-analysis pipeline.
The capabilities of the proposed method were presented on structures of engineering relevance, recorded by means of two shape measurement approaches: image-based reconstructions using photos taken with a hand-held device, and laser scanning.

\begin{figure}[!htbp]
  \centering
  \includegraphics[width=1.0\textwidth]{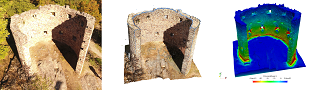}
  \caption{\textit{Tower ruin: photo taken by the UAV, point cloud and the maximum principal stresses computed by the FCM}}
  \label{fig:hocheppan}
\end{figure}

It is noted here that image-based algorithms do not exclusively work with pictures stemming from hand-held devices. 
Recent developments in the technology of UAV-s have made it possible to take inexpensive but high-quality pictures of objects of virtually any size.
In the field of cultural heritage preservation, this allows for an especially convenient way to assess the structural health of historical buildings.
As an example, Figure~\ref{fig:hocheppan} demonstrates a stress analysis conducted on the ruin of a medieval tower located at the ,,Hocheppan Castle'' in South Tyrol. 
The results were computed following the point cloud-based FCM analysis pipeline, starting from 120 pictures taken of the ruin using a DJI Phantom 4 drone.

While the presented approach is a first step towards establishing seamless connections between shape measurement techniques and numerical analyses, future research should address further questions concerning at least the following three aspects:
\begin{enumerate}
  \item The treatment of Dirichlet boundary conditions on non-conforming surfaces. For continuous surface descriptions, this can be resolved by applying these boundary conditions in the weak sense, using e.g. the penalty method or Nitsche's method -- allowing them to be formulated in terms of contour integrals that can be transformed into domain integrals, following the ideas presented in Section~\ref{ssec:pointCloudBC}. 
  \item Especially for historical masonries, the anisotropic behavior introduced by the interaction between the mortar and the building blocks needs to be modeled.
  \item Cracks, if present and visible in the original structure, should be incorporated in the corresponding numerical model. 
\end{enumerate}



\FloatBarrier
\bibliographystyle{ieeetr}
\bibliography{references} 


\end{document}